\documentclass[twocolumn,traditabstract]{aa}
\usepackage{graphicx}
\usepackage{amssymb,amsmath}
\def\kms{km\,s$^{-1}$}
\def\CO10{{\hbox {CO(1--0)}}}

\def\,{\thinspace}
\def\Msun{M$_\odot$}

\def\Lsun{L$_\odot$}

\def \Kkmspc{K\,\kms\,pc$^2$}

\def\cii    {\ifmmode{{\rm C}{\rm \small II}}\else{C\ts {\scriptsize II}}\fi}
\def\tex {\ifmmode{{T}_{\rm ex}}\else{$T_{\rm ex}$}\fi}
\def\tmb {\ifmmode{{T}_{\rm mb}}\else{$T_{\rm mb}$}\fi}
\def\tkin {\ifmmode{{T}_{\rm kin}}\else{$T_{\rm kin}$}\fi}

\begin{document}

\title{BR1202--0725: An Extreme Multiple Merger at $z$ = 4.7}

   \author{P. Salom\'e\inst{1}
           \and
           M. Gu\'elin\inst{2}
           \and
           D. Downes\inst{2}
           \and
           P. Cox\inst{2}
           \and
           S. Guilloteau\inst{3}
           \and
           A. Omont\inst{4}
           \and
           R. Gavazzi\inst{4}
           \and
           R. Neri\inst{2}
          }


   \institute{LERMA, Observatoire de Paris, 61 av. de l'Observatoire
     75014 Paris, France\\
              \email{philippe.salome@obspm.fr}
         \and
         Institut de Radio Astronomie Millim\'etrique (IRAM), Domaine
         Universitaire, 300 rue de la piscine, 38406 St Martin d'H\`eres, France
         \and
             LAB, Observatoire de Bordeaux, 2 rue de l'Observatoire,
         BP 89, F-33270 Floirac, France
         \and
             IAP, Institut d'Astrophysique de Paris, 98bis bd Arago,
         F-75014 Paris, France
             }

   \date{Received 5 July 2012 / accepted 27 July 2012}

\abstract{The radio-quiet quasar BR1202--0725 ($z=4.695$) is a
    remarkable source with a bright Northwest (NW) companion detected
    at submm and radio wavelengths but invisible in the
    optical. In the absence of amplification by gravitational lensing,
    BR1202--0725 would be the most luminous binary CO and far-IR source
    in the Universe.  In this
    paper, we report observations with the IRAM Plateau de Bure
    interferometer of BR1202--0725 in the redshifted emission of the
    CO(5--4) and (7--6) lines, the [C{\,\small I}]($^3$P$_2-^3$P$_1$)
    line, a high angular resolution ($0.3'' \times 0.8''$) 1.3\,mm map
    of the rest-frame, far-IR dust continuum, and a search for the
    CO(11--10) line.  We compare these results with recent ALMA data
    in the [C{\,\small II}] line.  Both the quasar host galaxy and its
    NW companion are spatially resolved in the molecular line emission
    and the dust continuum. The CO profile of the NW companion is very
    broad with a full width at half maximum of $\sim 1000 \pm
    130$\,\kms, compared to $\sim 360\pm40$\,\kms\ for the quasar host
    galaxy to the Southeast (SE).  The difference in linewidths and
    center velocities, and the absence of any lens candidate or
    arc-like structure in the field, at any wavelength, show that the
    obscured NW galaxy and the SE quasar host galaxy cannot be lensed
    images of the same object.  Instead, we find morphological and
    kinematic evidence for sub-structures in both the NW and SE
    sources.  We interpret these results as strong indications that
    the BR1202--0725 complex is a group of young, interacting, and
    highly active starburst galaxies.

}

\maketitle

\section{Introduction}

Submillimeter observations of galaxies and quasars at high redshift
provide invaluable clues about the activity of galaxy mergers
leading to the formation of massive galaxies, and about the
star-forming environment when the universe was young.
These clues come from the redshifted emission of dust heated by
newly-formed stars and from molecular and atomic lines
from the dense molecular gas in which the stars are born.
Since the early 1990s, massive galaxies at high
redshift have been observed with ground-based millimeter and submm
telescopes that can detect submm-selected starburst galaxies
and host galaxies of optically selected quasars. These detections are
pushing more and more into the epoch of re-ionization, and
are currently out to $z=7.1$; (Venemans et al.\ 2012).

\begin{figure*}
\includegraphics[width=8.6cm,angle=-0]{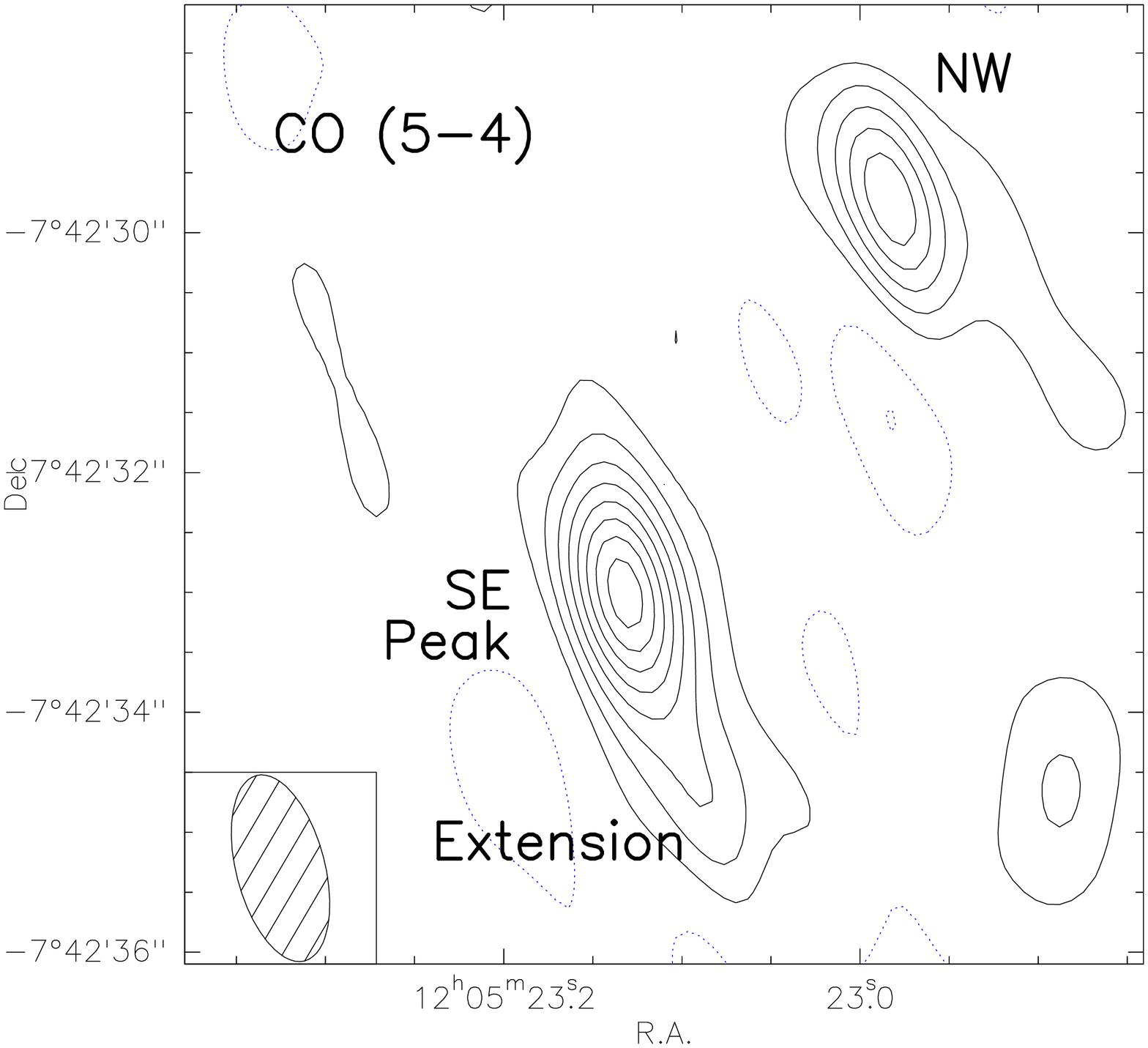}
\includegraphics[width=8.6cm,angle=-0]{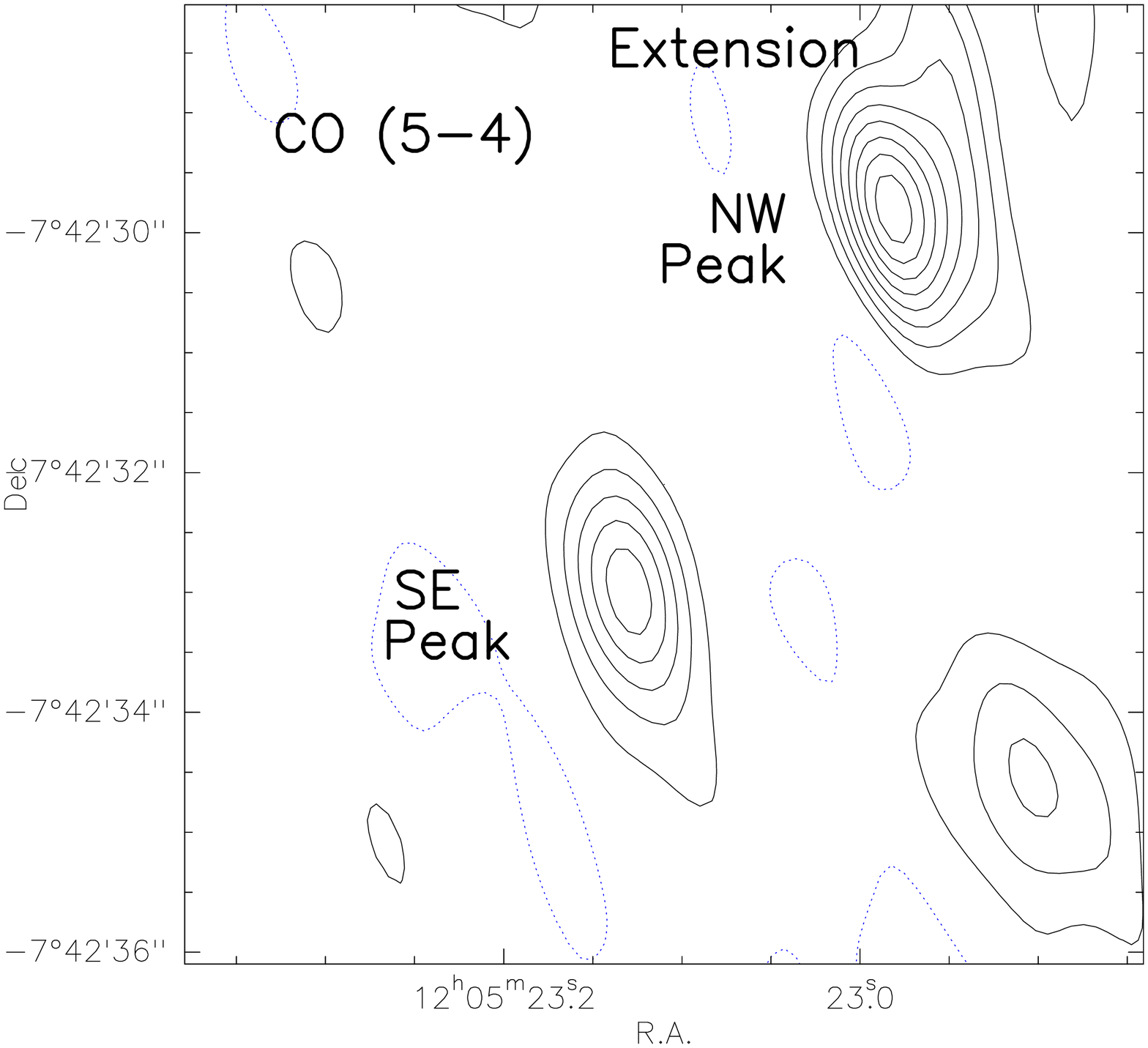}
\caption[CO54 uv-average maps of SE and NW.]{
Integrated CO(5--4) line maps, optimized to show total emission in the
SE and NW galaxies separately.  Source centroids and sizes are listed in
Table~1. Both maps have natural weighting with a
beam of $1.6'' \times 0.7''$ (PA 15$^{\circ}$; lower-left
box in left panel).
{\it Left:} CO(5--4) integrated over a 444\,\kms-wide band,
centered on the SE galaxy's line center velocity. Contour steps are
0.24\,mJy (1$\sigma$). The integrated CO(5--4) flux of the whole SE source,
including the extension to the SW, is 1.6\,Jy\,\kms .
{\it Right:} CO(5--4) integrated over a 1067\,\kms-wide
band, centered on the NW galaxy's line center velocity. Contour steps
are 0.15\,mJy (1$\sigma$) The integrated CO(5--4) flux of the whole NW
source, including the extension to the northwest, over the
1067\,\kms\ band, is  2.6\,Jy\,\kms\ (see Table~2)
  }
\label{fig:co54integ}
\end{figure*}

One of the first high-$z$ quasars that was detected and studied in
detail in both thermal dust emission and in CO lines is the very
luminous optically selected quasar BR1202--0725 at $z=4.7$ (McMahon et
al.\ 1994; Ohta et al.\ 1996; Omont et al.\ 1996; Benford et al.\
1999; Carilli et al. 2002; Iono et al.\ 2006;
Riechers et al.\ 2006).  The source is remarkable because it has two
distinct components: a Southeast (SE) source associated with the
optically luminous quasar and an obscured Northwest (NW) source with
no counterpart in the visible. The two submm dust continuum peaks,
separated by $3.7''$
(linear distance 24\,kpc),\footnote{ For linear sizes and luminosity
distances we took the standard cosmology with $H_{0}$ =
71\,\kms\,Mpc$^{-1}$, $\Omega_{M}$ = 0.27, $\Omega_{\Lambda}$ = 0.73
(Komatsu et al.\ 2011), and used the cosmological calculator by Wright
(2006).}  have nearly identical CO redshifts of $z=4.695$ (SE) and
 $z=4.693$ (NW).
The extreme infrared luminosity makes BR1202--0725 one of the
brightest high-$z$ sources ($L_{IR}$ = 3.7$\times 10^{14}$\,\Lsun,
Leipski et al.\ 2010, confirming the original estimate by McMahon et
al.\ 1994).  The far-IR part of the luminosity is $L_{\rm FIR}$ =
3.8$\times 10^{13}$\,\Lsun, with 1.2 and 2.6$\times 10^{13}$\,\Lsun\ for
the NW and SE galaxies (Iono et al.\ 2006).
The estimated star-formation rate in each component is $\geq \,
1000$\,M$_{\odot}$\,yr$^{-1}$  and the combined
molecular gas mass is $\rm \sim10^{11} \, M_{\odot}$
(Omont et al.\ 1996; Riechers et al.\ 2006).

The SE optical quasar host galaxy and the optically obscured NW source
are both detected in the CO $J= 1-0$ through $7-6$ rotational lines
(Ohta et al.\ 1996; Omont et al.\ 1996; Carilli et al.\ 2002; Riechers
et al.\ 2006) as well as in the [C{\,\small II}] emission line, the
main cooling line in the interstellar medium (Iono et al.\ 2006; Wagg
et al.\ 2012). At optical wavelengths, the SE component of
BR1202--0725 appears as a quasar with at least two faint companion
galaxies, seen in the optical (starlight) continuum (see our
Fig.~6).  The first of these optical continuum sources, found by
Fontana et al.\ (1996) and Hu et al.\ (1996), is centered 2.6$''$
northwest of the quasar, near to, but not coinciding with, the
obscured NW submm source, and it appears to be part of a filament of
optical starlight continuum emission extending over 4$''$ away from
the quasar.  Part of this optical continuum source is prominent in
Ly$\alpha$ emission at $z$ =4.702 (Hu et al.\ 1996; Petitjean et al.\
1996; Ohyama et al.\ 2004).  The second, fainter, optical starlight
continuum source is also a Ly$\alpha$ emission galaxy, located 3$''$
southwest of the quasar (Hu et al.\ 1997).

Up to now, detailed study of the profiles of the molecular and carbon
fine structure lines in BR1202--0725 have been hampered by lack of
sufficient bandwidth to cover the very broad molecular emission lines
of the NW galaxy (Omont et al.\ 1996; Carilli et al.\ 2002).  To
remedy this situation, we present in this paper new wide-bandwidth
spectral-line observations of CO and [C{\,\small I}] emission in
BR1202-0725, along with high-angular resolution observations of the
1.3\,mm thermal dust continuum.  These new data from the Plateau de
Bure interferometer (PdBI) are compared with submm data obtained with
ALMA that trace the [C{\,\small II}] emission line and the 0.9\,mm
thermal dust continuum emission (Wagg et al.\ 2012).  Our observations
improve significantly on previous studies and shed new light on the
kinematics of the SE and NW sources as well as on their
structures. These data indicate that BR1202--0725 is a very complex
group of merging starburst galaxies.

\section{Observations and data reduction}
The PdBI observed BR1202--0725 at 3 and 1.3\,mm in February and March
2007 and at 2\,mm in November 2011.  In winter-spring 2007, the 6
antennas were in the extended $A$ configuration with a longest
baseline of 760\,m. The receivers were successively tuned to 101.190
and 222.478\,GHz, close to the redshifted frequencies of the CO(5--4) and
CO(11--10) lines.  Due to the low declination
of the source, the synthesized beam was elliptical with a FWHM of
$1.6''\times 0.7''$ (PA 15$^\circ$) at 3\,mm,
and 0.85$'' \times 0.26''$ (PA 15$^\circ$) at 1.3\,mm.

The November 2011 observations were made in the intermediate $C$ configuration
at 142.6\,GHz, near the frequencies of
CO(7--6) (redshifted to 141.637\,GHz) and [C{\,\small I}]($^3$P$_2$--$^3$P$_1$)
(redshifted to 142.110\,GHz; hereafter, we refer to this neutral carbon line
simply as C{\,\small I}). The beam was
$3.0'' \times 1.8'' $ (PA 18$^\circ$) at 2\,mm.

\begin{figure} 
\centering
\includegraphics[width=8.6cm,angle=-0]{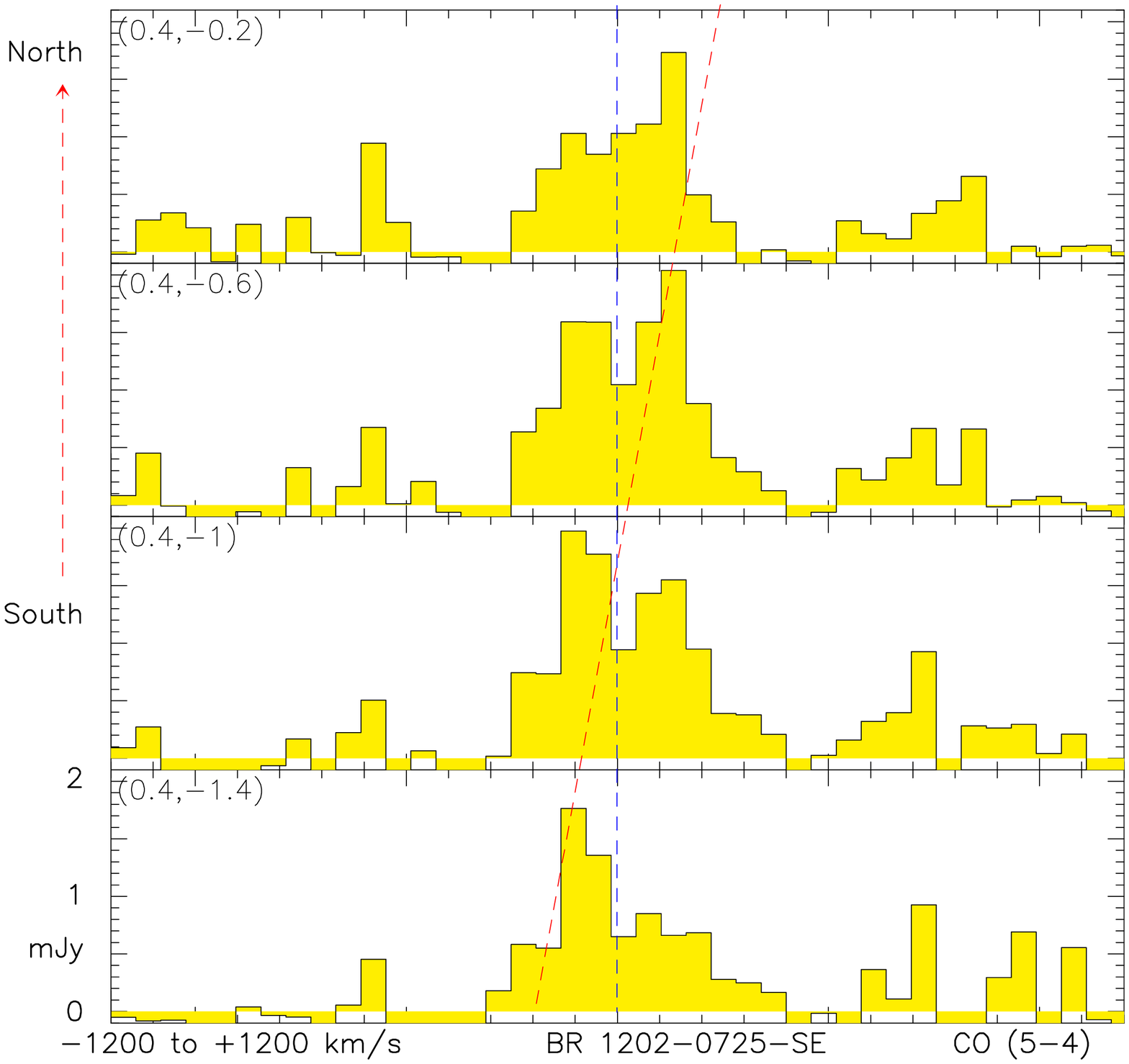}
\caption[CO54 spectra across SE]{CO(5--4) spectra across the SE galaxy,
in steps of 0.4$''$ in Dec.  The tilted dashed line indicates how
the two line components, at +110\,\kms\ in the north, and
 --70\,\kms\ in the south, cause the line centroid to shift
relative to zero velocity (dashed vertical line through center of the spectra),
in going from north to south (top to bottom).
 Position offsets in arcsec ($\Delta\alpha, \Delta\delta$, upper left
of each panel), are relative to the map phase center at
RA 12:05:23.11, Dec --07:42:32.10
(J2000).  Velocity offsets
(horizontal axes) run from --1200 to +1200\,\kms\ relative to 101.190\,GHz ($z$ = 4.6949).
The beam is $1.6'' \times 0.7''$ at PA 15$^\circ$.  Channel widths are 59.3\,\kms ;
1$\sigma$ = 0.48\,mJy; the data are not convolved with any smoothing function.
  }
\label{fig:SEspecgrid}
\end{figure}

\begin{figure}  
\includegraphics[width=8.6cm,angle=-0]{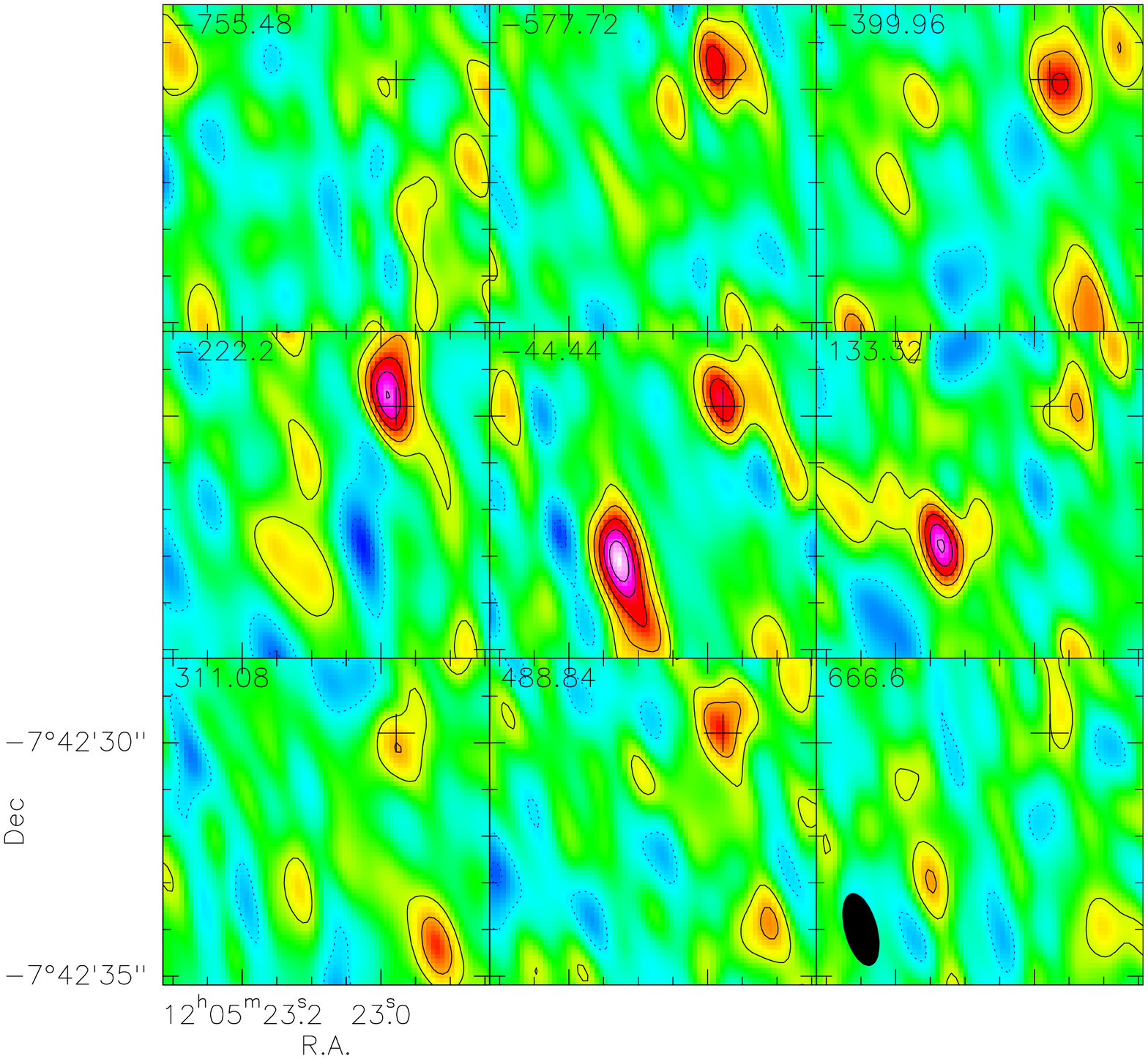}
\caption[CO54 channel maps]{CO(5--4) maps in
178\,\kms-wide channels.  The SE galaxy is strong at $-$44 and
+133\,\kms\ (middle row, center and right panels), while the NW galaxy
(upper right of each panel) has a main maximum at $-$222\,\kms\ (left
column, middle panel), and a weaker plateau extending to positive
velocities, notably in the panel at +489\,\kms\ (lower row, middle
panel).  Contour steps are 0.34\,mJy; 1$\sigma$ = 0.25\,mJy. The cross
in each box marks the position of the 1.3\,mm continuum from the NE
galaxy (Table~1).  Velocity offsets (upper left corner of each
box) are relative to 101.190\,GHz ($z$ = 4.6949).
The beam is 1.6$''\times 0.7''$ (black ellipse in bottom right panel).
}
\label{fig:NWchanmaps}
\end{figure}

The dual-polarization receivers operated in single sideband mode
(LSB), with SSB receiver noise temperatures of 40-50\,K. The
observations were done in dry weather, with system noise temperatures
of $\sim$140\,K at 2 and 3\,mm and 180\,K at 1.3\,mm, after correction
for atmospheric absorption and spillover losses. The IF bandwidth was
1\,GHz in 2007 (3000\,km\,s$^{-1}$ at 3\,mm, 1300\,km\,s$^{-1}$ at
1.3\,mm) and 3.6~GHz in 2011 (7600\,km\,s$^{-1}$ at 2\,mm). In the raw
spectra, channel spacings were 2.5\,MHz in 2007 and 2.0\,MHz in 2011.

Amplitude and phase were calibrated every 20 minutes on 3C~273, whose
flux was compared on each observing day to the continuum of the
ultra-compact H{\,\small II} region around the star MWC~349 (taken to
be 1.04, 1.48, and 1.67\,Jy at 101, 141 and 222\,GHz, respectively).

Pointing and focus were checked every 40 minutes. The PdBI ``seeing''
parameter, which is not a limit on the radio resolution, but an
indicator of the strength of the atmospheric phase fluctuations, was
0.33$''$ to 0.49$''$ at 3 and 2\,mm and 0.23$''$ to 0.26$''$ at
1.3\,mm. Short-term ($\rm < 1 min$) atmosphere phase fluctuations were
corrected with the aid of 3-channel monitoring receivers centered on
the 22\,GHz atmospheric water line.  Total on-source integration times,
excluding flagged data,
were 13.2\,h at 3\,mm, 5\,h at 2\,mm and 6\,h at 1.3\,mm. The data
were calibrated with the GILDAS CLIC program; both polarizations were
co-added and the maps were cleaned with the GILDAS MAPPING program.

\begin{figure*}
\includegraphics[width=8.7cm, angle=-0]{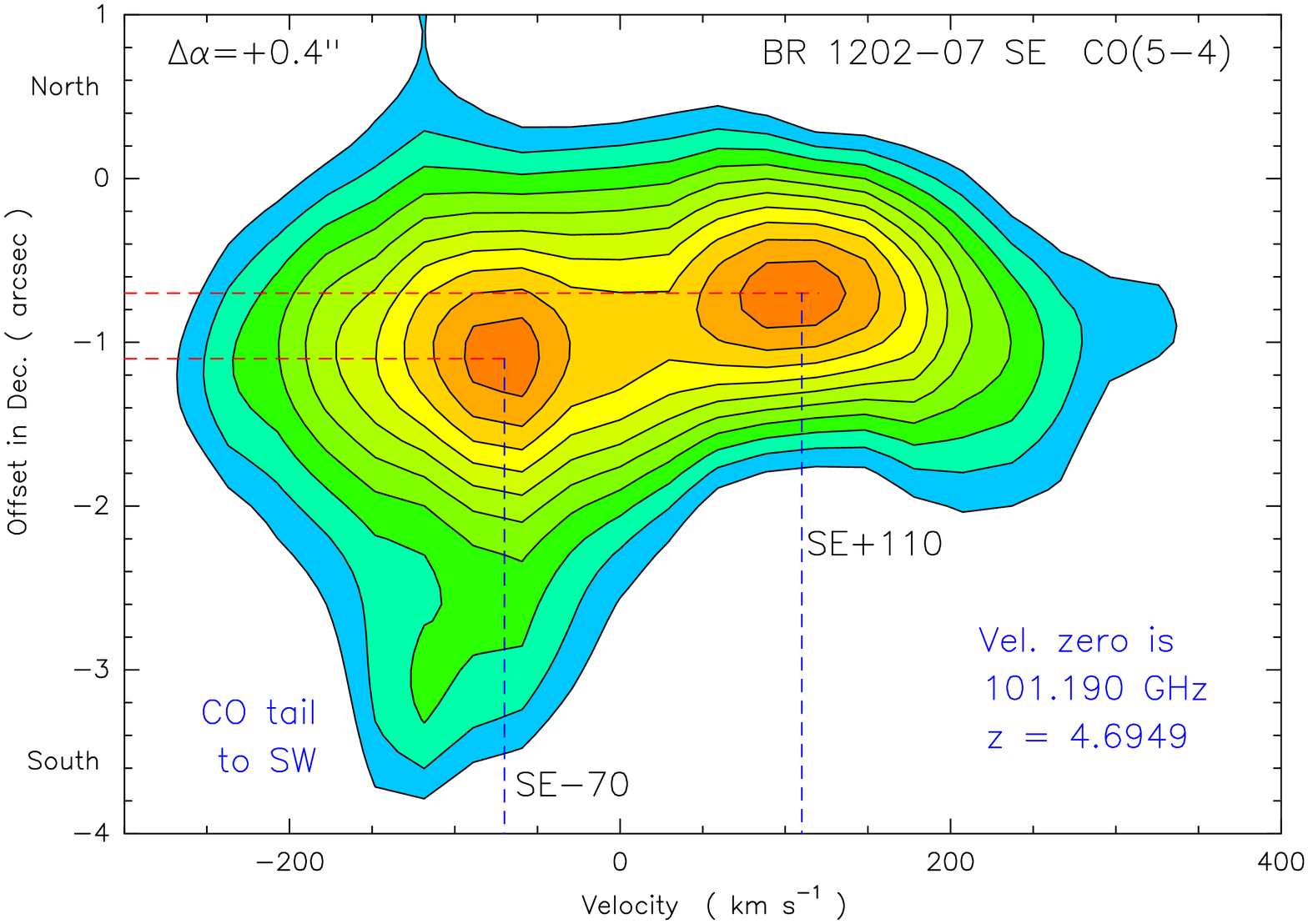}
\includegraphics[width=8.7cm, angle=-0]{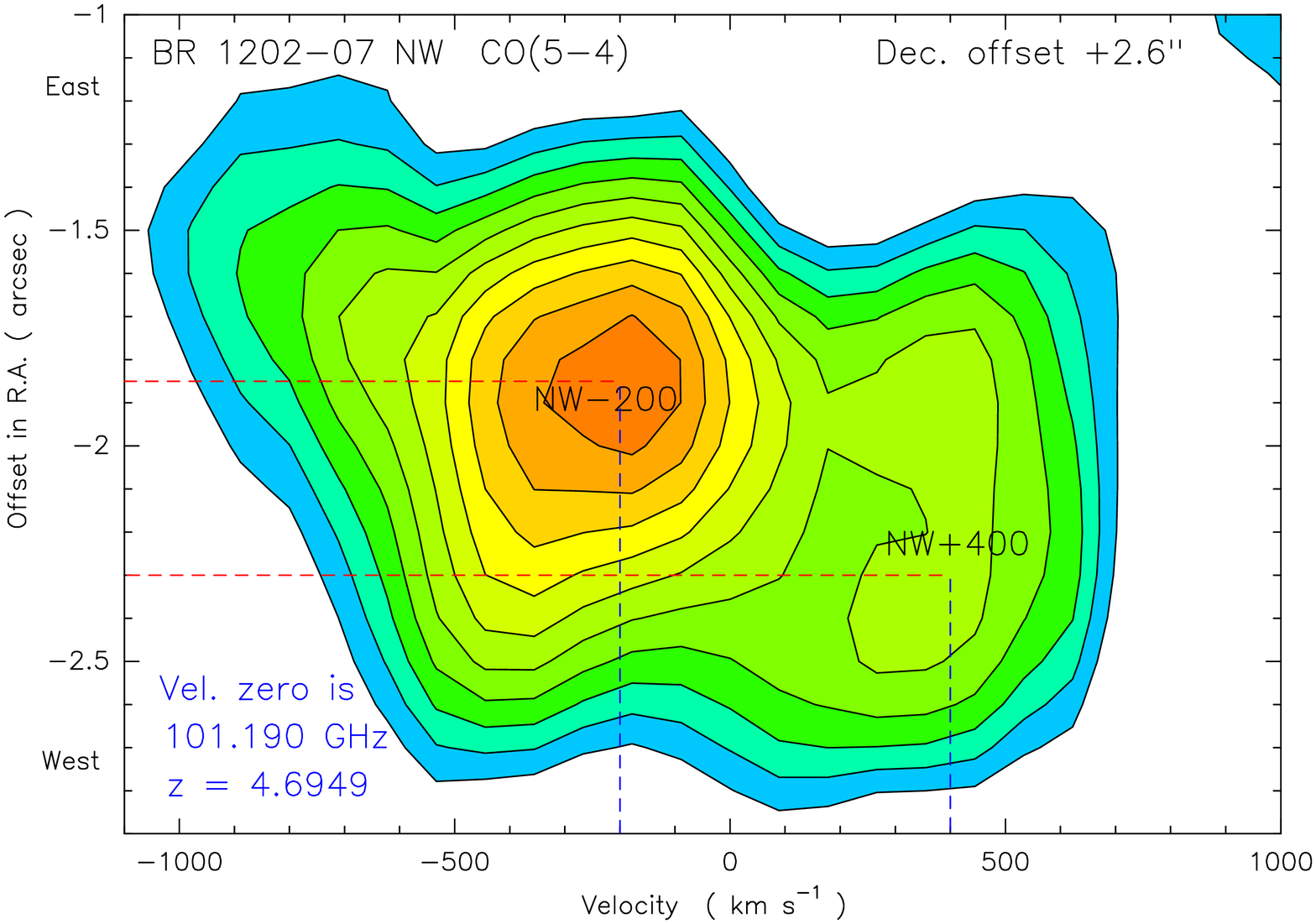}
\caption[P-V diagrams]{Position-velocity diagrams of CO(5--4) in BR1202--0725.
 {\it Left:} Dec.-velocity cut through the SE source.  The main line
 has two peaks, at $-$70 and $+$110\,\kms , separated by $+0.4''$ in
 dec.  At negative velocities, a narrow-line CO tail extends $>3''$
 southwest.  This tail to the SW is near, but does not coincide with,
 the weak SW continuum source found by Wagg et al.\ (2012).
 {\it Right:} R.A.-velocity cuts through the NW source. Note the large
 linewidth, which includes the strong line core near $-$200\,\kms, and
 the weaker plateau, that may be another merger galaxy, at +400\,\kms,
 and 0.6$''$ to the west or northwest.
 For both diagrams, velocity offsets are relative to 101.190\,GHz ($z$
 = 4.6949), and contours start at 2~$\sigma$ and increase in steps of
 1-$\sigma$ (0.37\,mJy/beam). The beam is $1.6'' \times 0.7''$ (PA
 $15^{\circ}$); position offsets are relative to the map phase center at
RA 12:05:23.11,  Dec --07:42:32.10 (J2000).
 }
\label{fig:channels}
\end{figure*}
\section{Results}
\subsection{The 3-mm data:  CO(5--4) line emission}
Figure~1 shows two different CO(5--4)
integrated intensity maps, made  separately
to best show the SE and NW galaxies,
with a limited velocity range for the
SE quasar host galaxy, and a much wider velocity range to show the
total emission from the  NW galaxy.
These maps indicate that the CO in the NW
galaxy is resolved to the north or northwest,
in the direction of the beam minor axis
(HPBW $0.7''$).  They also indicate a probable extension or
tail somewhat south or slightly southwest of the SE galaxy.

Figures 2, 3, and 5 show the CO(5--4) line data from the SE and
NW galaxies in detail not previously seen.  The most striking result
is how dissimilar the spectra are.  The CO profile of the SE galaxy, centered
at redshift $z= 4.6952$, has a FWHM linewidth of $\Delta V =
360\pm 40$\,km\,s$^{-1}$ (Table~2), comparable to
those of other high-$z$ quasar host galaxies detected in CO.
In the local universe, such a linewidth is typical of massive spirals seen
edge-on, and in the mid-range of linewidths for merging
galaxies like some of the low-$z$ ultraluminous infrared galaxies (ULIRGs), and the
high-$z$ submillimeter galaxies (SMGs; see, e.g., Fig.~6 of Bothwell et al.\ 2012).

In contrast, the NW galaxy has an extraordinarily broad profile with
$\Delta V_{\rm FWHM} \simeq 1000$\,km\,s$^{-1}$ and
$\Delta V_{\rm FWZP} \simeq 1700$\,km\,s$^{-1}$
(Fig.~4, right, and Table~2).
These values exceed the CO linewidths normally observed in quasars or
galaxies, with some notable exceptions like the 1500\,\kms\ -wide CO
line in the NGC~6240 merger (e.g., Feruglio et al.\ 2012), the
high-$z$ quasar-and-multiple-galaxy merger in SMM\,J02399-0136 (see
Ivison et al.\ 2010 and references therein), and the $z$ = 3.8
radio-galaxy merger in 4C41.17 (De~Breuck et al.\ 2005).  The main CO line peak
in the NW source is
blue-shifted by $-200$\,\kms\ relative to the CO centroid of the SE galaxy.
This main CO peak dominates
on the eastern side of the NW galaxy (Fig.~4, right).
Toward the western side of the NW galaxy,
there is a secondary spectral feature that is red-shifted
by +600\,\kms\ relative to the main CO peak in the NW galaxy.
These two spectral features at $-$200 and +400\,\kms\ suggest that the obscured NW
submm object is in reality two merging galaxies,
separated by about 0.6$''$ (4.0\,kpc).

The redshift of the +400\,\kms\ component happens to be the same as
the $z=4.702$ redshift of the Ly$\alpha$-1 galaxy between the NW and SE
galaxies (Hu et al.\ 1996; Fig.~6).  Like the CO lines
from the NW merger, the Ly$\alpha$ line is also very broad with
$\Delta V_{\rm FWZP} \simeq 1500$\,\kms\ and $\Delta V_{\rm FWHM}
\simeq 1100$\,\kms\ (Ohyama et al. 2004).  As in the CO in the NW and
SE sources, there are also multiple velocity components within the
main Ly$\alpha$-1 galaxy, over a range of 500\,\kms\ (Fontana et al.\
1998; Ohyama et al. 2004).  Hu et al.\ (1997) and Fontana et al.\
(1998) note that the C~{\small IV} line, seen in the quasar, is absent
in the Ly$\alpha$-1 galaxy, so the Ly$\alpha$ is not reprocessed quasar
radiation, but instead comes from star formation, at a rate of 15 to
54\,\Msun\,yr$^{-1}$.  The broad Ly$\alpha$ linewidths may thus be due
to superwind bubbles breaking out of the starburst and supernova
regions in their parent galaxies.  Because of these winds and
because of optical depth effects,  Ly$\alpha$
redshifts can give very biased measurements of the systemic redshifts
of galaxies, so the true redshift of the
Ly$\alpha$-1 object may differ from its measured Ly$\alpha$ redshift by
several hundred \kms .

As in the CO spectra (Fig.\,2), the CO position-velocity
diagram of the SE galaxy (Fig.~4, left) shows two main CO peaks
separated by 0.4$''$ to 0.5$''$ in position, and by 180\,\kms\ in velocity ,
that may indicate rotation of the
molecular disk in the inner $r$ = 2\,kpc of the quasar host galaxy.
There is also evidence of a weaker CO source or streamer extending
3$''$ southwest of the SE galaxy.  This tail is blueshifted by
--100\,\kms\ relative to zero velocity in Fig.\,4, left,
and it has a relatively
narrow linewidth (100 to 200\,\kms).
This weak CO tail to the southwest does not coincide, however, with
the 0.9\,mm continuum source reported by Wagg et al.\ (2012). Nor do
we see, at the position of their weak SW continuum source, any CO
emission at velocites corresponding to the edge of the bandwidth used
by Wagg et al., where they had proposed a tentative detection of
[C{\,\small II}].

\begin{figure*}
\includegraphics[width=18.0cm, angle=0]{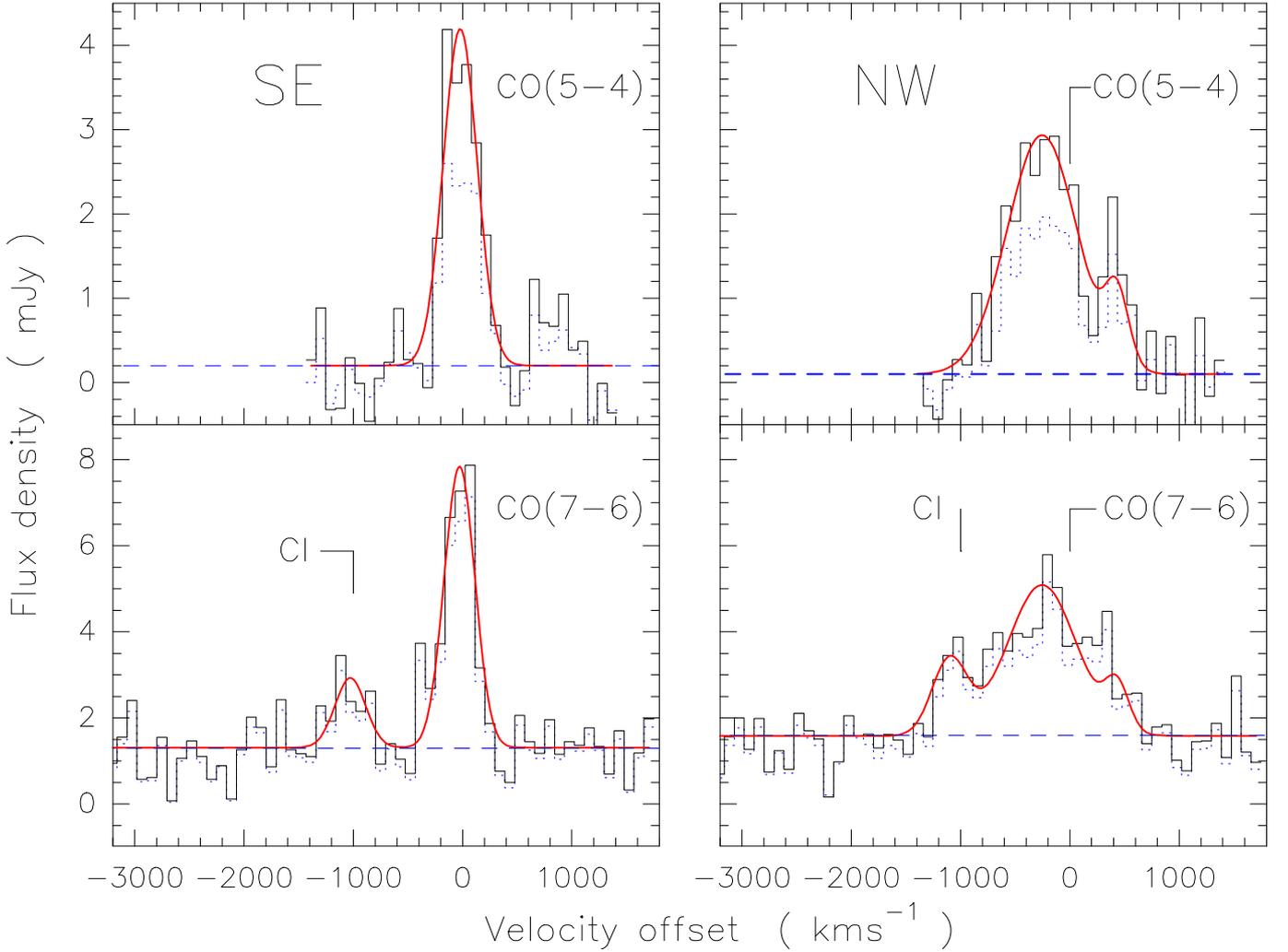}
\caption[CO54, CO76, and CI model-fit, spatially-integrated line spectra.]
{
CO(5--4), CO(7--6), and [C{\small I}]($^3$P$_2-^3$P$_1$) spectra
of the SE and NW components of BR1202-0725. Histograms
are spectra derived from fitting the interferometer visibilities by assuming
elliptical Gaussian sources with sizes (FWHM) of
1.0$''\times 0.6''$ (SE) and 0.8$''\times 0.6''$(NW).
Solid curves show Gaussian spectral-line fits to the profiles, and
dashed horizontal lines show continuum levels derived from these
Gaussian spectral fits.  The dotted histograms show alternative
model-fit spectra for assumed point sources.  The difference between
model-point-source and model-extended-source spectra is more striking
in CO(5--4) than in CO(7--6) because the larger beam in CO(7--6)
more easily includes all the flux in each of the SE and NW sources.
For the NW galaxy, two Gaussian velocity components
are fit to the CO(5--4) and CO(7--6) lines,
with a third Gaussian added to the CO(7--6) fit to include
the C{\small I} line.
Channel widths are 89\,\kms\  and 91\,\kms\  for
CO(5--4) and (7--6) respectively, and
velocity offsets (\kms ) are relative to CO(5--4) and (7--6) redshifted
to $z$ = 4.6952.
}
\end{figure*}


The position-velocity diagram of the NW galaxy (Fig.~4, right)
shows that the large linewidth in fact includes a strong core source
centered at $-$200\,\kms, and a weaker plateau, that may be another
merger galaxy, at +400\,\kms. This second line component, also seen in
the channel maps of Fig.\,3, is 0.6$''$ west or northwest
of the main source, in direction of the extension seen on the 1.3\,mm
dust continuum map.

The 3-mm dust continuum is not visible in channel maps or in
single-pixel spectra.  From averaging the line-free channels on both
ends of the spectra, we derive upper limits of 0.3\,mJy (2$\sigma$)
for both the SE and NW sources.
Spectral baseline fits to the
spatially-integrated spectra yield
a spatially-integrated, 3-mm continuum flux density of $0.2\pm 0.1$\,mJy
for the SE source, and are inconclusive for the NW source.

\begin{figure*} 
\centering
\includegraphics[width=7.9cm,angle=-0]{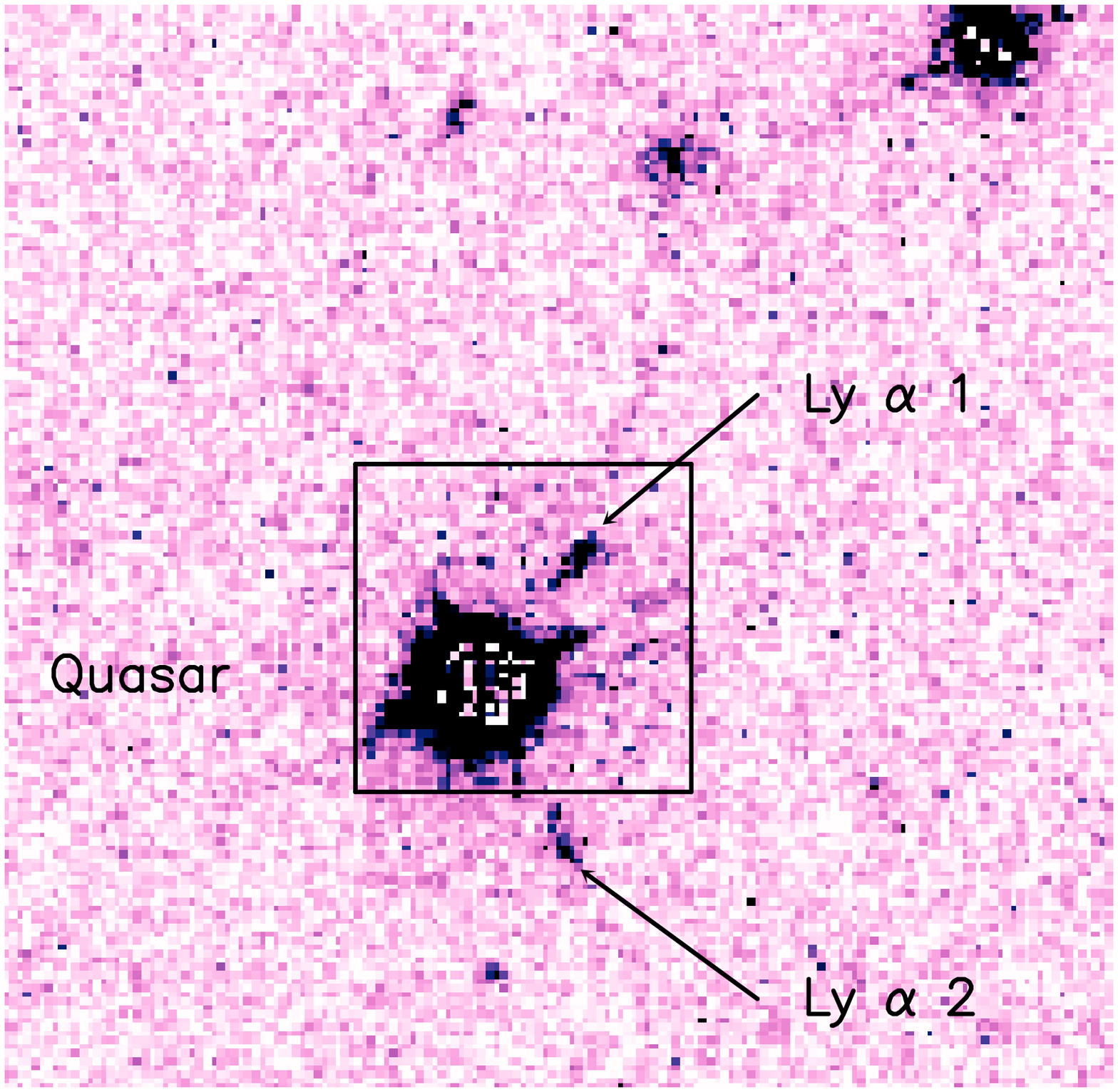}
\includegraphics[width=8.6cm,angle=-0]{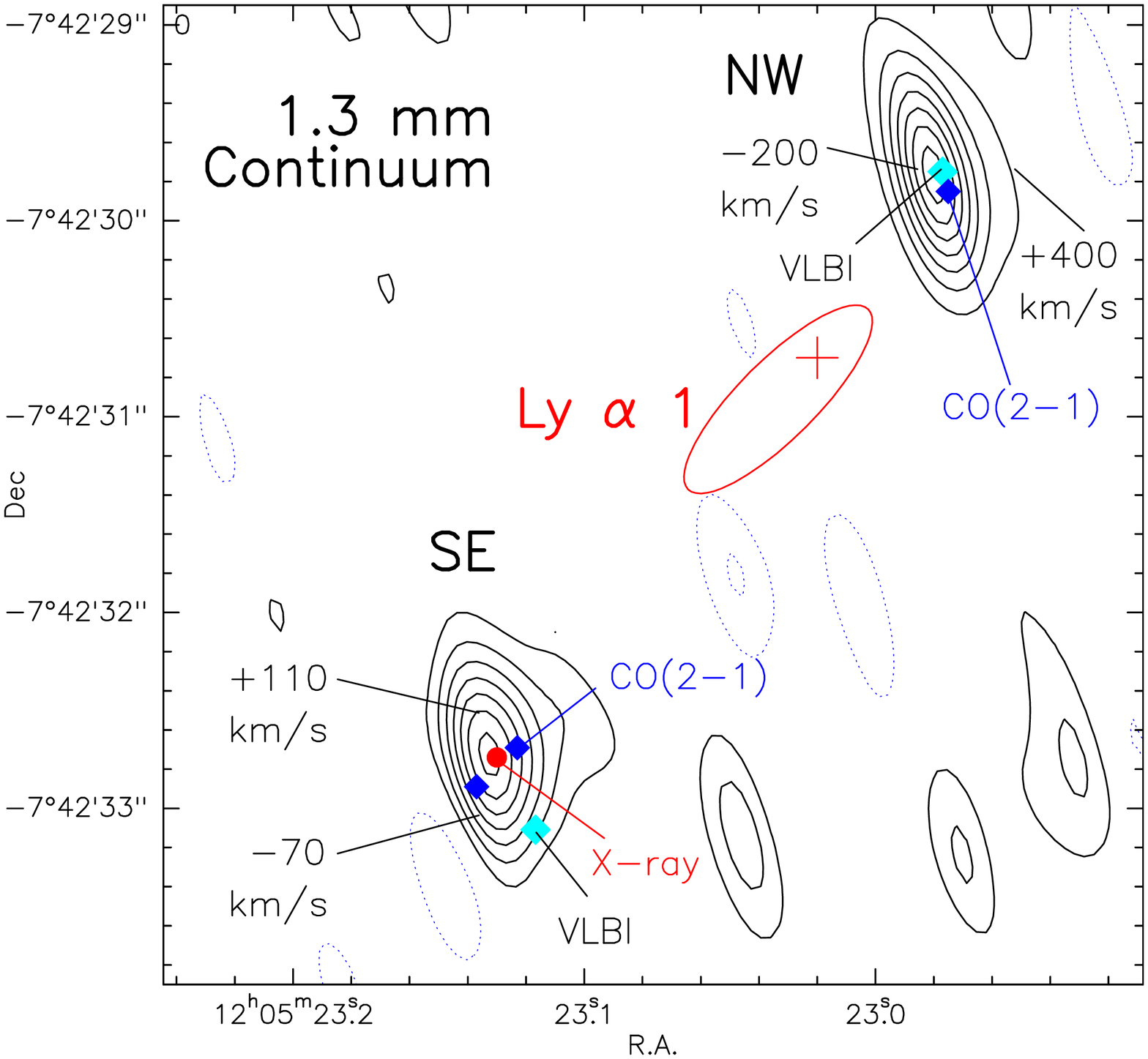}
\caption[HST+1.3\,mm dust continuum map]{{\it HST} image and 1.3\,mm
dust continuum map.
{\it Left:} {\it HST} F814W image of the BR1202--0725 region (Hu et
al.\ 1996) showing the quasar and the two galaxies with Ly-$\alpha$
redshifts measured by Hu et al.\ (1996; 1997).  In this filter, the
two Ly-$\alpha$ galaxies appear only in the starlight continuum, not
the Ly-$\alpha$ line.  The field is 16.4$'' \times$ 16.4$''$.  The
smaller box shows the 5$'' \times$5$''$ field of the 1.3\,mm continuum
map at the right.
{\it Right:} Contours show the BR1202-0725 dust continuum at 1.3\,mm
in steps of 0.26\,mJy/beam (1.5$\sigma$); negative contours are dashed,
beam = $0.85'' \times 0.26''$.
Next to the NW and SE galaxies
separated by 3.7$''$ are labels indicating the CO(5--4) velocity
components discussed in the text, with
offsets (\kms) relative to 101.190\,GHz
($z$ = 4.6949).
The Ly-$\alpha$-1 object,
near the NW galaxy, is indicated by a red ellipse.
The cross in the red ellipse marks the optical
continuum peak, 2.6$''$ from the quasar (Hu et al.\ 1996;
1997).  The Ly-$\alpha$ peak is near the southern part of the ellipse,
2.3$''$ from the quasar (Hu et al.\ 1996; Fontana et al.\ 1998;
Ohyama et al.\ 2004).
In the visible and near-IR, there is also a string of starlight continuum
emission extending over 4$''$ in the same direction, but displaced by 0.6$''$
from the  Ly-$\alpha$-1 object  (Hu et al.\ 1996;
Fontana et al.\ 1998; Ohyama et al. 2004). In the SE galaxy (lower left)
the central red dot indicates the
X-ray source, coincident with the quasar (Evans et al.\ 2010).  The
blue dots show the VLA CO(2--1) peaks on either side of the quasar,
and in the NW galaxy (Carilli et al.\ 2002).  The light-blue diamonds
show the extended non-thermal emission observed in VLBI at 1.4\,GHz
(Momjian et al.\ 2005).  Source positions are listed in
Table~1.
}
\label{fig:cont1mm}
\end{figure*}

\subsection{The 2-mm data:  CO(7--6) and C{\,\small I}  lines }
Observations at 2\,mm (142\,GHz) show that the CO and [C{\,\small I}]
lines are quite distinct in the SE source, where their spectral
separation is larger than their linewidths. In the NW source however,
due to their exceptionally broad widths, the CO(7--6) and C{\small
I} lines are blended together
(Fig.~5).  For the NW source, because of this blending of
the CO(7--6) and C{\small I}, and the two velocity components,
we fixed the CO(7--6) linewidth to be the same
as that of CO(5--4).  The results for CO(7--6) and C{\,\small I}, from
fitting a {\it single} Gaussian to each line are listed in
Table~2.  For the NW source, we show
an alternative fit in Fig.~5, upper right and lower right, with two
Gaussian velocity components in the NW source. In CO(5--4)
(Fig.~5, upper right), this yields a main
peak of 2.9$\pm 0.3$\,mJy, at $-$260$\pm 40$\,\kms, with a
linewidth of 737\,\kms\ (FWHM).  There is a secondary peak at +440$\pm
64$\,\kms , with a width of 268\,\kms.  The two
velocity components differ by 700$\pm 100$\,\kms,
which agrees within the errors with the separation of the two velocity
components in the channel maps of the NW source
(Fig.\,3), and in the position-velocity diagram
(Fig.\,4, right).  The CO(7--6) line of the NW source
(Fig.~5, lower right),
is blended with the [C{\,\small I}] line, and a three- or
four-component fit is needed.  In the SE source, where the CO and
[C{\,\small I}]
lines are well separated (Fig.~5, lower left), the
[C{\,\small I}] line has a width and center
velocity consistent with those of the CO(7--6) and CO(5--4) lines
(Table~2).

For the 2\,mm dust continuum, from both source maps and uv-fits,
 we obtain 2\,mm continuum flux densities of $1.3\pm0.2$ and
 $1.6\pm0.2$\,mJy for the SE and NW sources.  In the $3.0'' \times
 1.8''$ beam at 2\,mm, the dust continuum sources are unresolved.

\begin{table*}
\caption{Source positions and sizes.}
\begin{tabular}{l l l c c  c}
\hline
 		&R.A. 	  &Dec.		&$\Delta \theta$ from  	&Source	
								&
\\
{\bf Source} 	&12$^{\rm h}$05$^{\rm m}$  &$-07^\circ42'$	&SE-1.3mm 	&size   &Refs.
\\ 		
                &(J2000) &(J2000) &continuum	&(FWHM)
\\    			
\hline
\\
\multicolumn{5}{l}{ {\bf SE galaxy:} }
\\
SE 1.3\,mm continuum
	&23.129$^{\rm s} \pm 0.002^{\rm s}$   	
	&32.74$'' \pm 0.03''$ 		
	&{\bf 0.0$''$}
	&0.52$''\pm 0.06''$ 			
	&1  	
	\\
Chandra X-ray catalog
    &23.130$^{\rm s} \pm 0.02^{\rm s}$   	
	&32.74$'' \pm 0.3''$ 		
	&0.0$''$
	&--- 		
	&2
\\
CO(2--1) SE (nw subpeak)
	&23.122$^{\rm s} \pm 0.003^{\rm s}$
	&32.71$'' \pm 0.05''$
	&$0.1''$
	&$\leq 0.25''$
	&3
\\
CO(2--1) SE (se subpeak)
	&23.134$^{\rm s} \pm 0.003^{\rm s}$
	&32.92$'' \pm 0.05''$
	&$0.1''$
	&$\leq 0.25''$
	&3
\\
CO(5--4) SE centroid
    &23.130$^{\rm s} \pm 0.002^{\rm s}$   	
	&32.94$'' \pm 0.06''$ 		
	&0.2$''$
    &$0.5''\pm 0.1''$
	&1
\\
1.4\,GHz nonthermal SE source
    &23.1167$^{\rm s} \pm 0.04^{\rm s}$   	
	&33.109$'' \pm 0.3''$ 		
	&0.42$''$
	&0.30$''\times 0.17''$, 130$^\circ$ 			
	&4
\\
\\
\multicolumn{5}{l}{ {\bf NW galaxy:} }
\\
NW 1.3\,mm continuum
	&22.978$^{\rm s} \pm 0.002^{\rm s}$   	
	&29.79$'' \pm 0.03''$ 		
	&3.7$''$
	&0.26$''\pm 0.04''$	
	&1  	
\\
CO(2--1) NW
	&22.975$^{\rm s} \pm 0.003^{\rm s}$
	&29.85$'' \pm 0.05''$
	&$3.7''$
	&$\geq 0.5''$
	&3
\\
CO(5--4) NW centroid
&22.971$^{\rm s} \pm 0.004^{\rm s}$
	&29.67$'' \pm 0.06''$
	&$3.8''$
	&$0.7''\pm 0.1''$
	&1
\\
1.4\,GHz nonthermal NW source
    &22.977$^{\rm s} \pm 0.04^{\rm s}$   	
	&29.750$'' \pm 0.3''$ 		
	&3.7$''$
	&0.29$''\times 0.17''$, 115$^\circ$ 		
	&4
\\
\\

\hline
\multicolumn{6}{l}{{\it Position refs.}:
(1) This paper: our positions and sizes are from uv-fits to circular Gaussian models, and our
}
\\
\multicolumn{6}{l}{absolute astrometric accuracy is 0.1$''$;
}
\\
\multicolumn{6}{l}{
(2) Evans et al. (2010);
(3) Carilli et al. (2002); (4) Momjian et al. (2005).
}
\\
\\
\end{tabular}
\end{table*}

\subsection{The 1.3\,mm data: dust continuum and search for CO(11-10)}
The high angular resolution of 0.26$'' \times 0.85''$ (1.8$\times
5.6$\,kpc at $z$ = 4.7) attained with the extended configuration of
the PdBI at 1.3\,mm enables us to study the thermal
dust emission of BR1202--0725 in detail.

Figure~6 is a map of the 1.3\,mm (222\,GHz) continuum over the
whole 1\,GHz IF band. The band center was tuned to the
redshifted frequency of CO(11--10). This line was not detected in the
SE source, with a $3\sigma$ upper limit of 5\,mJy/beam
(Table~2).  We see a 1.7$\sigma$ hint of a line in the NW source,
but more data are needed to confirm it.

The dust continuum source positions and uv-fit sizes are listed in
Table~1.  The SE and NW sources are well separated in the
1.3\,mm continuum.  The peak positions of the dust emission are close
to the peaks of CO(5--4) reported in this paper, and also to the
CO(2--1) peaks (Carilli et al.\ 2002) and the non-thermal cm-radio
continuum peaks (Momjian et al.\ 2005).

The SE and NW sources are both resolved in the 1.3\,mm continuum. Our
model fits in the uv-plane yield
equivalent circular Gaussian FWHM diameters of
$=0.5''\pm 0.1''$ for the SE dust source,
and $0.4'' \pm 0.1''$ for the NW source.
More elaborate fits are not justified at present; a good rule of thumb
is that one needs S/N $\geq$ 10 to measure, e.g., major and minor axes and
position angles.
The spatially-integrated
 flux densities are $5.1 \pm 0.9$
(SE) and $4.0 \pm 1.3$~mJy (NW).

The weaker dust source to the southwest of the quasar, reported by
Wagg et al.\ (2012) at 340\,GHz, is not seen in the 1.3\,mm continuum map.
Its 0.9\,mm flux density of 1.9$\pm$ 0.3\,mJy implies a
1.3\,mm flux $\leq 0.5$\,mJy, too low (2$\sigma$) to be seen on our current map.

\section{Source properties}
In the Appendix to this paper, we review evidence that the
SE and NW sources in BR1202--0725 are not strongly gravitationally lensed.
All source properties given in the following sections therefore
have no correction factors for lensing.

\subsection{The northwest source}
The above results show that the NW source is
resolved in both the CO(5--4) line and the
1.3 mm continuum, as seen in Figs.\, 3, 5, and 6.
The CO line profiles clearly change,  between positions separated by
0.5$''$, on either side of the 1.3\,mm continuum
peak.  The
CO spectra and position-velocity diagrams across the NW source show
that the sub-sources, or  merging galaxies,
have velocities that differ by 600\,\kms, thereby
explaining the unusually large linewidth of the overall
CO profile.

\subsection{The southeast source}
As noted above, the CO(5--4) spectra of the SE galaxy show two line
peaks separated by 180\,\kms, and 0.4$''$ to 0.5$''$ (Figs.\,2
and 4).  These two {\it velocity} peaks are probably the same as the
two {\it spatial} peaks of the SE galaxy, already seen by Carilli et
al.\ (2002) who used a VLA beam of $0.3''\times 0.2''$ to resolve the
CO(2--1) emission into two components, oriented SE--NW and separated
by 0.3$''$. These two components are on either side of the quasar
(Fig.\,6), and may be in opposite sides of a disk rotating
around the nucleus.  In the 1.3\,mm dust emission, the quasar host
galaxy also appears to have a weaker extension 0.5$''$ to the
northwest (projected distance 3.3~kpc; Fig.~6).

From VLBI observations with the VLBA, the VLA, and the
 GBT,  Momjian et al.\ (2005)
 claim that the SE 1.4\,GHz nonthermal radio continuum peak is
apparently 0.4$''$ south of the quasar (Fig.~6, right, and Table~1),
while the VLBI NW source coincides with
the 1.3\,mm NW dust peak.
The SE displacement might be further evidence that the SE and NW
sources have multiple components.  The synchrotron emission sources
seen in VLBI at 1.4\,GHz  are
extended (diameters 0.3$''$, or 2\,kpc), with intrinsic brightness
temperatures of $2\times 10^4$\,K at their 8\,GHz rest frequency. With
a typical nonthermal $T_b$ spectral index of $-$2.7, they would have a
rest-frame 1.4\,GHz $T_b$ of 2$\times 10^6$\,K --- about 500 times higher
than in the central hundred parsecs of M82.  This non-thermal radio
brightness is comparable with those of compact starbursts in ULIRGs.
The BR1202-0725 nonthermal radio sources are therefore
also likely to be the products of extreme starbursts
rather than active galactic nuclei (AGNs; Momjian et al.\ 2005).

 \begin{table*}
      \caption[Line and continuum observed parameters.]{Observed parameters of CO and atomic lines, and continuum fluxes.}
         \label{tab:parameters}
          \begin{center}
           \begin{tabular}{lccccccc}
            \hline
Line   &Reference   &Line               &Line            &Center              &Integrated            &Line luminosity         &Continuum
\\
and    &frequency   &peak$^{\rm a}$    & width$^{\rm a}$ &velocity$^{\rm a}$  &intensity$^{\rm b}$   &$L^\prime$\,$^{\rm b}$
                                                                                                           &$S_{\rm cont}$\,$^{\rm b}$
\\
Source &GHz   &mJy                 &\kms             &\kms                &Jy\,\kms              &10$^{10}$\,\Kkmspc         &mJy
\\
\hline
\smallskip
{\bf CO(5--4)} & 101.185  \\
SE            & --- &$4.1\pm 0.4$ &$363\pm 37$   &\  $-17\pm 17$ &$1.6\pm 0.2$  &$5.3\pm 0.8$  &$0.2\pm 0.1$ \\
\\
NW-main peak  & --- &$2.9\pm 0.3$ &$737\pm 90$   &$-260\pm 40$ &$2.3\pm 0.3$  &$7.8\pm 1.2$  &$<0.3$ (2$\sigma$) \\
NW-extension  & --- &$0.9\pm 0.1$ &$268\pm 50$   &$+440\pm 64$ &$0.3\pm 0.1$  &$0.9\pm 0.1$  &$<0.3$ (2$\sigma$) \\
NW-total      & --- &---          & ---          &---          &$2.6\pm 0.4$  &$8.7\pm 0.1$  &$<0.3$ (2$\sigma$) \\
\hline
\smallskip
{\bf CO(7--6)}  & 141.637  \\
SE & ---  &$6.7\pm 0.5$           &$316\pm 40$  &\  $-27\pm 13$ &$2.3\pm 0.2$ &$3.9\pm 0.4$    &$1.3\pm 0.1$      \\
\\
NW-main peak  & --- &$3.6\pm 0.3$ &$737\pm 90$   &$-260$  &$2.8\pm 0.3$ &$4.8\pm 0.5$    &$1.6\pm 0.1$       \\
NW-extension  & --- &$1.2\pm 0.1$ &$268\pm 50$   &$+440$  &$0.3\pm 0.1$ &$0.9\pm 0.1$    &--- \\
NW-total      & --- &---          & ---          &---     &$3.1\pm 0.4$ &$5.7\pm 0.6$    &$1.6\pm 0.1$  \\
\hline
\smallskip
{\bf [C{\,\small I}]}($^3$P$_2$--$^3$P$_1$) & 142.109  \\
SE & ---  &$1.7\pm 0.4$  &316 &---   &$0.6\pm 0.2$   &$0.9\pm 0.3$    &$1.3\pm 0.1$ \\
NW & ---  &$1.8\pm 0.3$  &368 &---   &$0.7\pm 0.2$   &$1.2\pm 0.4$    &$1.6\pm 0.1$ \\
\hline
\smallskip
{\bf CO(11--10)} & 222.471\\
SE &---  &$<3.8$ (2$\sigma$)    &---  &---  &$<1.4$ (2$\sigma$)    &$<1.0$ (2$\sigma$)    &$5.1\pm 0.9$ \\
NW &---  &$2.7\pm 1.7$          &---  &---  &$2.9\pm 1.8$          &$2.0\pm 1.2$          &$4.0\pm 1.3$       \\
\hline
\smallskip
 {\bf [C{\,\small II}]}$^{\rm c}$ & 334.96 \\
SE &---  &27        &$328\pm 6$       &210   &$9.6\pm 1.5$    &$3.0\pm 0.5$    &$18\pm 3$ \\
NW &---  &20        &$722\pm 12$    & 60   &$14.7\pm 2.2$   &$4.6\pm 0.7$    &$19\pm 3$ \\
\hline
\hline
\smallskip
{\bf CO(2--1)}    & 40.479  \\
SE &---  &$0.77\pm 0.1$ &--- &---  &$0.23\pm 0.04$  &$4.8\pm 1.0$    &$<0.16$ (2$\sigma$)    \\
NW$^{\rm d}$ &---  &$0.66\pm 0.1$ &--- &---  &$0.39\pm 0.08$  &$8.1\pm 2.0$    &$<0.16$ (2$\sigma$)     \\
\hline
\end{tabular}
\end{center}
\begin{list}{}{}
\item[] $^{\rm a}\,$Line peak fluxes, FWHMs, and  velocities at the source peaks,
  from single-Gaussian spectral fits in velocity.
\item[] $^{\rm b}\,$Line intensities, $L^\prime$, and $S_{\rm cont}$ are
   over 1.0$''\times 0.6''$ (SE) and 0.8$''\times 0.6''$(NW);
line luminosity  $L^{\prime}$ from formula of Solomon et al.\ (1992).
\item[] $^{\rm c}\,$[C{\,\small II}] data from Wagg et al.\ (2012).
\item[] $^{\rm d}\,$CO(2--1) flux in NW source (Carilli et al.\ 2002)
corrected for limited VLA bandwidth by multiplying by 1.5,
  the ratio of the CO(5--4) flux in the full line width,
  to that in the limited VLA band of --525 to 175\,\kms .
\end{list}
\end{table*}

\subsection{Molecular gas properties}
We estimated the molecular H$_2$ gas densities and
kinetic temperatures in the NW and SE sources by comparing the
observed CO luminosities with single-component model brightness
temperatures in the RADEX 1-D escape probability program (Van der Tak
et al.\ 2007).    The data on
CO(5--4),(7--6), and (11-10) are from
this paper; the CO(2--1) intensities were derived from Carilli et al.\
(2002) (see Table~2).  These are the only data sets that have both
good sensitivity and clear separation of the SE and NW sources, and
they thus provide the first opportunity to model the CO Spectral
Energy Distribution (CO-SED) for the NW and SE sources separately
(Fig.~7).

We have no data from high density tracers like HCN and not yet
enough spatial resolution for CO lines above CO(5--4). The
kinetic temperatures and H$_2$ densities are therefore poorly defined.
There are two main constraints: (1) in both sources, the CO(5--4) and (7--6)
line luminosities, $L^\prime$, are nearly equal, and (2) in both sources,
the CO(11--10) is very weak, and not detected, at least in the SE source,
with a luminosity at least four to five times
lower than the CO(5--4) luminosity (Table~2).

If the gas temperature is in the range 40 to 50\,K, i.e., typical of
gas temperatures in ULIRGs, and close to the dust temperatures derived
below, these constraints imply that the optical depths of the CO lines
are high, at least up to the CO(7--6) line.  Thanks to the collisional
excitation and the resonant radiative trapping, the populations of levels
$J < 8$ are then nearly thermalized ($T_{ex} \simeq T_k$), so that
line brightness temperatures and luminosities $L^\prime$ remain constant.
In higher rotational transitions, the line opacity decreases, and unless the
H$_2$ density is unrealistically large ($>10^6$\,cm$^{-3}$), this leads to
a steep decrease in line brightness temperature.

The key is the high optical depth in CO.  The practical consequence
for the BR1202-0725 sources is that
in flux units, because of the weighting by $\nu^2$,
the CO spectral-line energy distribution peaks around $J$=8
(Fig.~7).

For comparison with escape probability models, we assumed the CO
abundance was $X_{\rm CO}$ = 10$^{-4}$, and we took the
microturbulence to be 1\,\kms\,pc$^{-1}$, both of which are standard
values often assumed for Galactic molecular clouds.  We took the background
radiation field to be the  cosmic background, at a
temperature of 15.6\,K at $z$=4.7, and we assumed gas
kinetic temperatures of 43\,K, the value estimated from the continuum dust
spectra.  In this zone of kinetic temperature,
escape probability models can reproduce the two constraints
listed above for a nearly constant value of the product
\begin{equation}
n({\rm H}_2)\cdot {N_{\rm CO} \over \Delta v } = 1.2 \times 10^{23}\ ,
\end{equation}
where $n({\rm H}_2)$ is the H$_2$ density in cm$^{-3}$, and
$N_{\rm CO} / \Delta v$ is the CO column density, in cm$^{-2}$ per
microturbulent linewidth $\Delta v$, in \kms .

Realistic, optically-thick combinations would thus be
$n$(H$_2$) = 2$\times 10^3$\,cm$^{-3}$,
and  $N_{\rm CO} / \Delta v$ =  $6\times
10^{19}$\,cm$^{-2}$\,(\kms )$^{-1}$,  for both the SE and NW galaxies.
The upper range would be
$n$(H$_2$) = 2$\times 10^4$\,cm$^{-3}$, and $N_{\rm CO} / \Delta v$
= $6\times 10^{18}$\,cm$^{-2}$\,(\kms )$^{-1}$.
A solution in this range would be
 consistent with the gas mass (see next section), and
with the dust becoming optically thick near rest-frame 100\,$\mu$m.

  Lowering the CO abundance, or CO
column density, or increasing the microturbulence would require higher
densities of molecular hydrogen to fit all the data.  These simple
models predict a CO(5--4) brightness temperature of 25\,K, i.e., close
to the Rayleigh-Jeans equivalent of ($T_{ex}-T_{bg}$), the difference
between the (Planck) excitation temperature and the cosmic background
at the CO(5--4) rest frequency of 576\,GHz.  Figures 2 and 3
show that the peak CO(5--4) brightnesses are 2.0 and 1.5\,mJy
beam$^{-1}$ in the SE and NW galaxies respectively.  At 101.2 GHz, the
1.6$''\times 0.7''$ beam yields 107\,K\,Jy$^{-1}$, so in our beam, the
observed peak brightness temperatures, corrected for (1+$z$), are 1.2
and 0.9\,K for the SE and NW galaxies. If the source sizes at the
velocities of the CO spectral peaks are about 0.3$''$, or about
the measured sizes of the dust sources, or about half the size of
the entire line sources integrated over all velocities, then the true,
deconvolved CO(5--4) peak line brightness temperatures, corrected for
beam dilution, are 17 and 13\,K for the SE and NW galaxies, or within
a factor two of the predicted brightness temperature of 25\,K from the
simplest radiative transfer model.

More complicated fits with
multi-temperature and multi-density components would change these results,
and
meaningful values for $T_{k}$ and $n$(H$_2$)
are hard to derive because of the parameter
degeneracy.  In spite of the uncertainties,
we can conclude, in agreement with previous studies,
that much of the molecular gas in both of
the BR1202--0725 sources is dense and warm.

\begin{figure}
{\includegraphics[width=8.9cm,angle=-0]{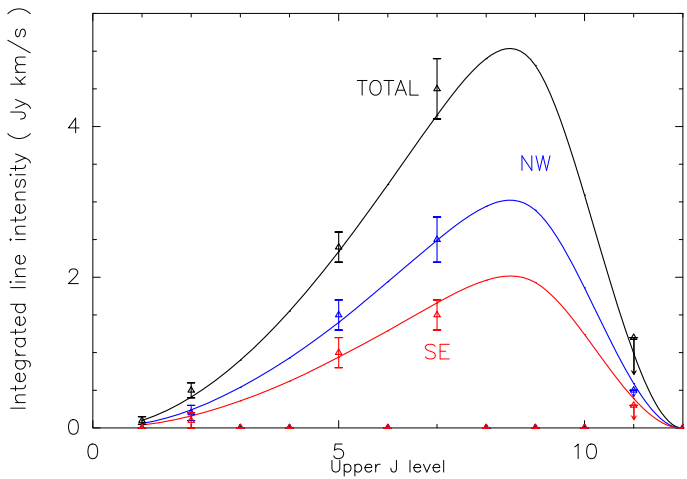}}
\caption[CO SEDs for NE and SW sources.]{
CO line peak flux densities vs.\ upper $J$ level,
for the SE (red lower curve) and NW (blue middle curve) galaxies of
BR1202-0725.
The top black curve shows their sum.  CO(1--0) data are from
Riechers et al. (2006); CO(2--1) from Carilli et al. (2002);
all other data are from this paper.
The curves correspond to the radiative-transfer models described
in the text.
  }
\label{fig:co-sed}
\end{figure}

\subsection{Gas mass and dynamical mass}
We use the CO(1--0) line luminosity-to-(H$\rm_2$+He)-gas mass
conversion factor derived for the center of ULIRGs by Downes \&
Solomon (1998),
\begin{equation}
M_{\rm gas}/L'_{\rm CO_{1-0}}=0.8 M_\odot {\rm (K\ km\ s^{-1}\ pc^2)^{-1}}
\label{eqMass}
\end{equation}
to estimate the gas mass in the SE and NW galaxies.

We took the CO(1--0) luminosities to be equal to the CO(5--4)
luminosities.  There are two reasons to do this, observational
and theoretical.  Firstly, direct measurements show that the CO(5--4)
luminosities are equal to the CO(2--1) luminosities (Table~2),
which means that in the CO levels $J<5$, the CO is optically thick and
thermalized.  Furthermore, the total (SE+NW) CO(5--4) luminosity of
1.3$\times 10^{11}$\,\Kkmspc\ is not only the same as the total
(SE+NW) CO(2--1) luminosity, but also the same within the errors as
the total (SE+NW) CO(1--0) luminosity of 1.0$\times 10^{11}$\,\Kkmspc\
obtained in the single-dish CO(1--0) observations by Riechers et al.\
(2006).

Secondly, at $z=4.7$, the CO(1--0) excitation temperature and the gas
kinetic temperature have to be warmer than the cosmic microwave
background temperature of nearly 16\,K for CO(1--0) to be in emission;
otherwise CO(1--0) would be in absorption (up to now never observed in
high-$z$ sources), or have too low a contrast relative to the CMB to
be detectable.  In standard radiative transfer models for gas warmer
than 20\,K (in order to have CO(1--0) in emission), and with $n(H_2)$ greater
than 500\,cm$^{-3}$ (to collisionally excite the CO), and with high
enough $N_{\rm CO}$ (to make the CO opaque enough to be detectable),
the brightness temperatures of CO(1--0) and CO(5--4) are the same. Hence
at $z=4.7$, at every place you detect CO(1--0), you will also detect
CO(5--4) --- with the same brightness.  At $z=4.7$, there is no extra
``reservoir" of ``cold" CO(1--0) that is currently detectable.

We therefore take the CO(1--0) luminosities equal to the CO(5--4)
luminosities, namely $L^\prime$ = (5.3 and 7.8)$\times
10^{10}$\,\Kkmspc\ for the SE and NW-main galaxies (Table~2),
and thus derive $M_{\rm gas}$ = (4.2 and 6.2)$\times 10^{10}$\,\Msun\
for SE and NW-main, respectively.  For both galaxies, the CO
luminosities and consequently, the gas masses, are larger than in
local ULIRGs, and comparable to the mean values of the SMGs
listed by Bothwell et al.\ (2012; see their Fig.~6).  For BR1202-07,
these high CO luminosities and gas masses are probably partly the
result of the sources' multiplicity.

Similarly, we can estimate the gas mass from the CI line,
using the formulae of Weiss et al.\ (2003; 2005).  Assuming
the same excitation temperature as for CO, and a [CI/H$_2$] ratio of
3$\times 10^{-5}$ by number (1.8$\times 10^{-4}$ by mass), we derive
gas masses of  (5.3 and 6.7)$\times 10^{10}$\,\Msun\
for the SE and NW-main galaxies, i.e., similar to the gas mass
estimated from CO.

Lower limits to the dynamical masses can be estimated from the CO
linewidths and the source sizes.  For the inner parts of the sources
(the sizes derived from the fits to the 1.3\,mm continuum), we
derive the dynamical masses of the SE and NW sources from

\begin{equation}
M_{\rm dyn}(\leq R)   = { {R\cdot V^2} \over G}  =232\ R\cdot V^2_{\rm rot}
\end{equation}
where $1/G$ = 232, when $M_{\rm dyn}$ is in \Msun, $R$ is in pc, and
$V_{\rm rot}$ is in \kms\ (e.g., Solomon et al.\ 1987). In principle,
the rotation velocity could be estimated from the observed
linewidths $\Delta V_{\rm FWHM}$, which would
at face value imply true rotation
velocities of 360 and 970\,\kms\ for the SE and NW sources.
In practice, these objects are dominated by the gas mass rather than the
stellar mass, and have thick disks, with chaotic, merger-driven,
and probably non-circular
gas orbits.  Above all,
the turbulent velocities in the molecular gas in advanced mergers and ULIRGs
are very large , of the order of
100 to 150\,\kms\  (Downes \& Solomon 1998).  Furthermore, the
position-velocity diagram for the NW source (Fig.~4b) indicates
that the NW CO source is multiple; the strong core of the
CO line, with a width of $\sim 400$\,\kms\ corresponds to the main NW
galaxy and its 1.3\,mm dust continuum, while the plateau in the CO
emission to higher velocities may be the other galaxy in the NW merger,
possibly the weaker satellite extension seen on the dust continuum
map.  If we take the 400\,\kms\ CO linewidth associated with the
0.6$''$-diameter CO source, and correct downward for turbulent broadening,
the true rotation velocity would be 300 to
350\,\kms, comparable to a rotation velocity we might deduce for the SE
galaxy.
Hence the dynamical mass within a radius of 2\,kpc ($\sim
0.3''$) in both galaxies would be of the order of $6\times 10^{10}$\,\Msun\
for each of the SE and NW sources.

Comparison of these dynamical masses of $6\times 10^{10}$\,\Msun\
with the gas masses  of $4.2 \times
10^{10}$ and $6.2 \times 10^{10}$\,\Msun\  indicates that in the inner
parts of the SE and NW galaxies, most of the mass is in the form of
molecular gas.  It also means that it would be difficult in
these galaxies to have a CO-to-H$_2$ mass conversion factor much
larger than the value for ULIRGs; otherwise, the gas mass
within the central $r$ = 2\,kpc would
greatly exceed the dynamical mass.

\begin{figure}
{\includegraphics[width=8.9cm,angle=-0]{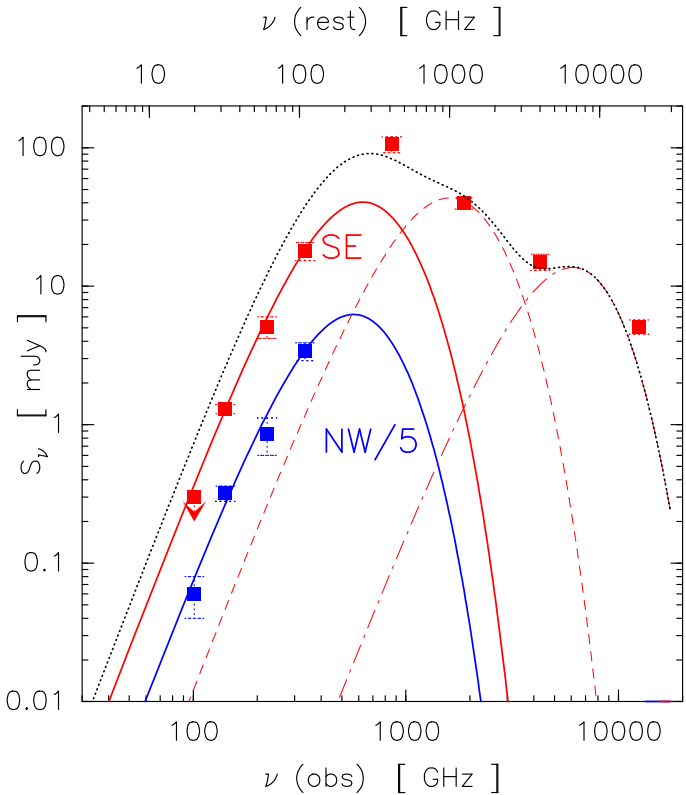}}
\caption[Dust continuum spectra for NE and SW sources.]{
Dust continuum spectra for NE and SW galaxies in BR1202-0725.  Flux
densities from 100 to 340\,GHz are from the mm/submm interferometer
data, as these are the only observations that spatially separate the
NW and SE galaxies in the cool dust continuum: 101, 142, and 222\,GHz
from this paper, 340\,GHz from Wagg et al.\ (2012).  Because the
mm/submm dust continuum fluxes of the two galaxies are nearly the
same, we plot the NW galaxy fluxes (blue lower curve) divided by five, for
better readbility.  Data points at 350, 160, 70, and 24\,$\mu$m are
from Benford et al. (1999), Leipski et al.\ (2010), and Hines et al.\
(2006).  The red curves show model dust components for the SE galaxy: a
starburst-heated 50 K component and two quasar-heated warm (130 K) and
hot (410 K) dust components. The uppermost dotted curve is the total
emission.
  }
\label{fig:dust spectra}
\end{figure}

\subsection{Dust mass and  infrared luminosity}
The dust properties in BR1202--0725 have been discussed by Benford et
al.\ (1999), Iono et al.\ (2006), Leipski et al.\ (2010) and Wagg et
al.\ (2012).  To these studies, the results in this paper now add new
measurements at 1.35, 2, and 3\,mm that update previous data from
Omont et al.\ (1996) and Guilloteau (2001).  For the NW and SE
sources, the updated dust continuum fluxes are given in Table~2.

For an estimate of the dust mass, we assumed a source size
of 0.5$''$, for consistency with the uv-fit sizes derived from the observed
1.3\,mm continuum sources, and a dust mass opacity coefficient of
$\kappa(\nu_r) = 10\cdot\nu_r^\beta$, where $\nu_r$ is the rest-frame
(emitted) frequency in THz, $\kappa$ is in cm$^{-2}$ gm$^{-1}$ of
dust, and we took the index $\beta$ = 2.  This absorption coefficient
agrees with extensive literature estimates for very dense molecular
clouds with coagulated dust particles (e.g., Kr\"ugel \& Siebenmorgen 1994;
Ossenkopf \& Henning 1994).
Because our observing frequencies correspond to rest
frequencies well into the far-IR, where the large quantity of dust
starts to become opaque, it is important in these estimates of dust
mass not to use the standard formula for optically-thin dust.
Instead, one must do a proper fit of the dust source function minus
the background source function (see eq.(1) of Weiss et al.\ 2007).
Because at $z$ = 4.7, the cosmic background temperature is ~16\,K, it
is important to subtract the background from the dust source function,
as it reduces the source-to-background contrast for the observed dust
flux.  Applying this method to the mm/submm interferometric data only,
which are the only observations that spatially separate the
dust continuum of the NW and SE galaxies, we can fit
the observed data (Fig.~8) with
a dust temperatures of 40 to 50\,K in both sources.
The luminosities of these ``cool", ``starburst" dust components alone,
from integrating over the range 20 to 1000\,$\mu$m (rest-frame)
under the modified Planck functions in our models, are $6.3 \times
10^{12}$\,\Lsun\ (SE) and $\rm 1.2 \times 10^{13}$\,\Lsun\ (NW),
comparable with values derived in previous studies.
These values yield dust masses of
$\sim 4\times 10^8$\,\Msun\ in both the NW and SE
sources.

The flux densities from 350 to 24\,$\mu$m (rest-frame 60 to
4.2\,$\mu$m), follow the usual quasar-template monotonic decrease (a
flat curve when plotted as $\nu S_\nu$), due to quasar-heated dust
from a continuous distribution of dust layers of increasing density
and decreasing mass and size as one moves inward toward the quasar.
Following the model of Weiss et al.\ (2003) for the Cloverleaf quasar,
we approximate this continuous distribution with just two components,
a ``warm" (130\,K) layer of size 0.11$''$, and a ``hot" (410\,K) layer
of size 0.014$''$.  This approximation (Fig.~8) shows that nearly all
the rest-frame mid-IR and near-IR flux comes only from the SE galaxy
(the quasar); the mid-to-near-IR contribution of the NW galaxy is
negligible.  The 350\,$\mu$m flux measured by Benford et al.\ (1999;
rest frame 60\,$\mu$m), comprises not only the fluxes of the SE and NW
``starburst" components, but also a substantial contribution (30 to
50 percent) from the quasar warm dust component.  The {\it Herschel}
PACS fluxes at 160 and 70\,$\mu$m (Leipski et al.\ 2010; rest frame 28
and 12\,$\mu$m) are entirely from the quasar-heated ``warm" and ``hot"
dust components; the contributions from the SE and NW ``starburst"
components are negligible.  The {\it Spitzer} MIPS flux at 24\,$\mu$m
(Hines et al.\ 2006; rest-frame 4.2\,$\mu$m) is mainly from the
quasar-heated ``hot" dust component.  There is no contribution from
the SE and NW ``starburst" components.

\subsection{Star-forming activity and merging}
From the far-infrared luminosities derived above for the ``cool", ``starburst"dust components,
one can estimate
the star formation rates following Kennicutt (1998):
\begin{equation}
\centering
{\rm SFR\ \ [M_\odot/yr]} = 1.7 \times 10^{-10} L_{\rm{FIR}}\ \ {\rm [L_\odot]}
\end{equation}
which yields 1000 and 2000\,\Msun\,yr$^{-1}$ for the SE and NW
components of BR1202-0725.  The Kennicutt SFR relation is for a
Salpeter Initial Mass Function (IMF).  Adopting other IMFs, such as
done by Kroupa (2001) or Chabrier (2005), would reduce these SFRs by
about a factor of two.  Even with a more top-heavy IMF however, it is
clear that both the SE and NW galaxies have very intense starbursts,
in multiple active regions.  Depending on the IMF, the minimum times
to consume all the molecular gas are 40 to 80\,Myr (SE) and 30 to
60\,Myr (NW), but longer if there are several, spaced-out,
million-year bursts.  These times are comparable to the
$\sim$40\,Myr-consumption times in typical submm galaxies (SMGs) at
$z\sim 2$ (Tacconi et al.\ 2008).  The derived gas masses in
BR1202-0725 are also about the same as in typical SMGs at $z\sim 2$.
It is possible, that the quasar in the SE galaxy, and possibly a
radio-quiet, dust-obscured AGN in the NW galaxy, contribute to the
dust heating, and the star formation rate is overestimated. The other
possibility is that we are observing an advanced multiple merger in
BR1202-0725 close to the peak in its starburst luminosity curve, as
shown by the wealth of components in the BR1202-0725 merger, including
possibly two sources within each of the SE and NW galaxies, the active
quasar, and the companion Lyman $\alpha$ protogalaxies. This
wealth of activity may at least in part be a signature of the
evolution of massive, high-redshift galaxies.
Similar complexity will probably be discovered in the
SMGs at $z\sim 2$, once they have been studied in comparable detail to
 BR1202-0725.

Rather than a merger of just two gas-rich galaxies, familiar from many
 ULIRGs at low redshift, the complexity of the BR1202-0725
 interaction, with multiple galaxies or pre-galactic building blocks, may
 itself be a characteristic of the assembly of massive galaxies at
 high redshifts.  Another high-redshift example of a complex merger is
 the $z$ = 3.8 merger in the radio galaxy 4C41.17, where there are at
 least two CO sources, with a total CO(4--3) line width of 1000\,\kms\
 around the powerful AGN source of the extended radio jets (De~Breuck
 et al.\ 2005), as well as numerous near-IR companions (Graham et
 al.\ 1994), that appear like raisins in a pudding contracting to form
 a very massive galaxy.
At least one of the companion regions,
4C41.17-South, has been shown to have the same redshift as
the main AGN galaxy, 4C41.17-North
(Van Breugel et al. 1999).
Another relevant example is the $z$=2.8
 multiple-galaxy merger in SMM\,J02399-0136 (see Ivison et al.\ 2010
 and references therein), which also shows ~1000\,\kms-wide CO lines
 coming from the merger of a far-IR luminous but highly obscured
 starburst (their L2SW galaxy, coincident with a Ly$\alpha$ cloud), a
 BAL quasar host galaxy (their L1 galaxy) and two other components in
 the field (the L2 and L1NW galaxies).  All four of these
objects are at the same redshift (Ivison et al.\ 2010), and must be
kinematically associated, not just seen in projection.
As in BR1202-0725, the ensemble
 of merger components in SMM\,J02399-0136 form a vast and complex
 nursery, not only for new stars, but also for the formation of the
 future massive galaxy itself.

In the new data, the NW galaxy and the SE quasar host galaxy
each break up into sub-components. Based on
the present work, the BR1202--0725 group consists of at least four
distinct sources (six when counting the faint Ly$\alpha$
companions). The proximity and similar redshifts of these galaxies or
pre-galactic objects suggest that the SE quasar host galaxy, the
obscured NW submm galaxy, and the Ly$\alpha$ objects are all in a
gravitationally-bound group of galaxies that have decoupled from the
cosmological expansion, and are probably merging together.  All of
these objects thus have exactly the same cosmological redshift, and
the slight differences in their measured redshifts are due to true
kinematic motions due to their mutual interaction within their "local
group."

In the standard cosmology, a $z$ = 4.7 galaxy is at a co-moving
(tape-measure) distance from us of 7.6 \,Gpc, and the lookback time is
12.4\,Gy.  When the galaxy emitted the light we now receive, its
recession velocity was 2.3\,$c$, the Hubble parameter, $H(z$ = 4.7)
was 505\,\kms\,Mpc$^{-1}$, the only relevant component of the universe
was matter, with $\Omega_M$ = 0.98, and dark energy was irrelevant,
with $\Omega_\Lambda$ = 0.014 (see e.g., formulae in Ryden 2002;
Liddle 2003).  In the usual interpretation, the BR1202-0725 group is
in the potential well of a overdensity of matter relative to the mean
density of the universe, that, long before the $z$ = 4.7 epoch, had
decoupled from the cosmic expansion, so the galaxies in the group have
only true kinematic relative motions governed by the total mass in the
overdensity region.  The greater neighborhood of the BR1202-0725
overdensity region as a whole may be similar to the numerous
``redshift sheets" seen at high-$z$ in the Hubble Deep Field (Cohen et
al.\ 2000), or to the recently-found galaxy groups at $z$ = 5.2 (Capak
et al.\ 2011; Walter et al.\ 2012).  In interpretations of the
same-redshift groups, (e.g., Cohen, Hogg, \& Blandford 2000, their
sect.\ 6), these regions extend over several Mpc and have virial
masses of a few times 10$^{13}$ to 10$^{14}$\,\Msun, resulting in true
kinematic velocity dispersions (not cosmological redshifts) of a few
hundred \kms .  Our measured line-of-sight velocity differences of the
different BR1202-0725 objects are of this order, and due to the close
approaches (projected separations 3 to 25\,kpc) in the individual
mergers, the relative velocities can be even higher.

In other respects, the BR1202-0725 objects agree with the general
picture of high-redshift protogalaxies.  Numerous deep-field
morphological studies show that at high-$z$, there are no spirals (e.g.,
Driver et al.\ 1998; Abraham \& van den Bergh 2002).  The
galaxies are gas-rich, puffed-up disks or spheroids,
and much smaller than modern-day galaxies.  The
data of, e.g., Oesch et al.\ (2010) show that the half-light radii
decrease as (1+$z$) to the 1.1 to 1.3 power.  At $z$ = 4.7, typical
galaxy radii are 700 to 1400\,pc, depending on the mass.  The high end
of this range is close to sizes measured in the present paper for the
BR1202-0725 SE and NW galaxies.

\section{Conclusions}
Our high-resolution millimeter observations of the BR1202-0725 galaxy
complex at $z=4.7$, in both line and continuum emission, reveal new
aspects of this extreme multiple merger. Morphological and
dynamical evidence together with the lack of any lens candidate or
arc-like structure in the field, at any wavelength, definitely rule
out the hypothesis that the obscured NW galaxy and the
SE quasar host galaxy are lensed multiple images
of the same background object. Instead, the present data
indicate that the
images of BR1202--0725 directly show a group of merging and very
active starburst galaxies (Fig.~6, right).

The SE galaxy merger contains at least two sources. The
main SE core source coincides with the optical quasar's position and
shows internal kinematics compatible with rotation and a CO line
profile (FWZP= 700 km/s) typical of a massive disk galaxy.
To the southwest, the CO(5--4) line is narrower,
with a linewidth of 200\,\kms, and
blueshifted by $-$180\,\kms\ relative to the main SE velocity peak.
This fainter CO does not coincide in
position with the weaker SW continuum source reported in the ALMA
data (Wagg et al. 2012), which may be a third, fainter, galaxy in the merger.

The NW galaxy contains at least two sources, a main core
source, and a fainter companion to the west/northwest at a velocity
different by 600\,\kms\ from the main NW core.  This second galaxy in
the interaction accounts for the unusually broad total CO linewidth
(FWZP $>$ 1500 km/s) of the NW galaxy merger.

The far-IR luminosities of both sources are high, of order 10$^{13}$\,\Lsun,
indicating tremendous star-forming
activity, with estimated star formation rates greater than
1000 \,\Msun\,yr$^{-1}$. Since BR1202--0725 is radio quiet, its
activity must be dominated by the massive starbursts that occur
simultaneously in the galaxies interacting in the NW and SE
components.  We are obviously witnessing an extreme merging event in a
group of galaxies, probably the youngest complex merger so far
known. Future higher angular resolution observations of BR1202--0725
will allow us to better disentangle the properties of the individual
starbursts that are interacting and merging, resulting in the
remarkable appearance of this high-redshift object.

\begin{acknowledgements}
These observations were done with the IRAM Plateau de Bure
Interferometer. IRAM is supported by INSU/CNRS (France), MPG (Germany)
and IGN (Spain).  The authors are grateful to the IRAM staff for
their support. We thank Emmanuel Momjian for information on the VLBI
positions, Juan Uson for useful comments, and the referee for very helpful
criticism.
\end{acknowledgements}



\appendix
\section{Absence of gravitational lensing}
Whether the submm/radio companion is a distinct object or
    a gravitational image of the quasar host galaxy is an important question
    because in the absence of magnification, BR1202--0725 would be the
    most luminous binary CO and far-IR source in the Universe.
The observational facts do indeed speak against gravitational lensing.
The difference in linewidths of the NW and SE sources, the absence of
optical or UV emission from the NW source, and the multiple CO and
dust components in each source, are all difficult to reconcile with a
gravitational lensing scenario to account for the appearance of
BR1202--0725. In the following discussion of gravitational lensing, we
definitely rule out the hypothesis that the SE and NW sources are
multiple lensed images of the same background source.

\subsection{Different linewidths.}
The main piece of evidence comes from the greatly different
molecular and atomic emission line profiles observed towards the NW
and SE sources.  These NW and SE CO and carbon-line profiles are so
different in their widths and peak velocities, that they cannot be
gravitational images of a single object.  Differences
in velocities and linewidths might
occur in multiple lensed images of optical
lines, if, for example, a multiply-imaged quasar were locally magnified by a
micro-lensing event that would affect only one of the images.
Such an effect  cannot affect the submm CO and
carbon lines, however, because these low brightness,
high luminosity lines are emitted over a large ($\geq 2$-kpc), relatively cool
(50 to 100\,K) region, and unlike the small, hot, accretion disk of a
quasar, cannot be affected by micro-lensing.
Differential image magnification between the NW and SE CO sources
can also be ruled out; although the SE and NW CO sources have quite different linewidths,
their dust continuum fluxes are nearly the same.
The dust sources and the molecular line sources are observed at the same wavelengths,
and their measured sizes ($\sim 0.5''$) are roughly the same, so there is no evidence for any
differential magnification between the SE and NW sources at submm wavelengths.

\subsection{No optical counterpart of the NW galaxy.}
The second piece of evidence comes from the dissimilarity of the
optical and mm/submm images of the quasar and its main companion, the
NW source.  The SE submm dust and molecule
component is associated with an optically-bright
quasar whose light is not seen at the position of the NW submm component.
The nearly unsurmountable problem is to make a gravitational
lensing model that would produce a single-spot optical image
and a double-spot submm image.  To first order, gravitational lensing
is achromatic {\it if the source size is the same at different
wavelengths}.

 But the quasar optical light (rest-frame UV) and its
broad lines come from the black hole accretion disk on a scale of
10$^{-2}$\,pc, while the low-brightness-temperature CO lines and submm
dust emission must be of kpc-scale size to be detectable at all at
high redshifts.  The molecular disk area is thus 10$^{10}$ times
larger than the quasar accretion disk area, and this can lead to
dramatic differences in lensing amplification and lensed image shapes
between optical and submm wavelengths.

Two examples are IRAS F10214+4724, and APM 08279+5255.  In F10214, the
quasar is too obscured to be visible, but the AGN optical narrow-line region
is magnified by a factor of 100, while the much larger submm molecular disk
is magnified by only a factor of 10.  (see models by Broadhurst \&
Leh\'ar 1995; Trentham 1995; Eisenhardt et al.\ 1996).  In APM 08279,
the quasar's rest-frame UV through mid-IR is also magnified by factor of 100,
while the cold part of the submm molecular disk is magnified by much less
(see model by Egami et al.\ 2000, their Fig.~9).  But even in these
models with large differences in magnification between the
optical and submm, it is nearly impossible to simultaneously produce a
double image at submm wavelengths, and only a single image at optical
wavelengths.

In a lensing scenario, one would have to imagine a galaxy on the line
of sight somehow producing a double image only of the molecular disk,
but not of the optical quasar, and then, in addition to this first lens that
images only the molecular disk, a second lens producing a strong
micro-lensing of the quasar to
make an optically bright, single, quasar image, with no micro-lensing
amplification of the much larger molecular disk.  This
overly-complicated scenario would not work at near-IR
wavelengths of 3.6 and 4.6$\mu$m, (rest
wavelengths 6300 and 8070\,\AA), however, because
the emitting region, the hot outer parts of
the circumnuclear ring around the quasar, are large compared to a
typical micro-lensing Einstein radius (e.g., Sluse et al.\ 2011).

Contradicting this scheme, the {\it Spitzer} archive images at 3.6
and 4.6$\mu$m show that the quasar is indeed much brighter ($>$100 times)
than any object toward the NW, in the near-IR.  In fact, the light
that is seen in the NW in {\it Spitzer} data is actually the extended
starlight continuum from the Ly$\alpha$~1 companion galaxy, which is
also near-IR bright in the $R$, $I$, and $K$ bands (see discussion of their
{\it Spitzer} data by Hines et al.\ 2006).  The submm dust and molecular
disks in the SE and NW galaxies that are discussed in the present paper
cannot be detected in the {\it
Spitzer} near-IR images, because they are too cold; their flux is
hopelessly far down the Wien side of the blackbody curve, and they do
not radiate at all at rest-frame optical wavelengths (Fig.~8).

All these difficulties to find a lensing model that would produce a single-spot
optical quasar image and a double-spot submm image, such as a galaxy-mass
lens for a double-image molecular disk only, simultaneously with stellar-mass
micro-lensing for the quasar only, that doesn't really work for the
near-IR images, lead us to conclude that a lensing explanation is a
blind alley.

\subsection{Absence of a candidate deflector to make a double image.}
Deep optical images from the {\it HST} and VLT do not show any
evidence of a massive foreground object capable of deflecting the
light of the quasar of BR1202--0725 into two widely-separated
images. Nor do they show any evidence of any arc-like structure,
neither on galaxy-lensing angular scales, nor on cluster-lensing
angular scales.

Because the SE and NW sources have similar fluxes in the submm
continuum and in the CO, [C{\,\small I}], and [C{\,\small II}] emission
lines, the most natural hypothesis for gravitational lensing would be
a fairly circularly symmetric mass distribution located close to the
mid-point of the $4\arcsec$ line connecting the two sources. Deep
ESO-NTT imaging (Fontana et al.\ 2000) of the system along with {\it
BVrIK} data (Giallongo et al.\ 1998) allow us to put tight upper
limits on the deflector brightness.

To produce an image splitting of $\Delta \theta =4\arcsec$, a
deflector at redshift $z_L$, that for simplicity we approximate as a
Singular Isothermal Sphere, would need to have a velocity dispersion
given by:
\begin{equation}
  \sigma= 186.25\ {\rm km/s}
  \cdot  \sqrt{ \frac{\Delta \theta}{2 w}}\,,
\end{equation}
where $w=d_{LS}/d_S$ is the ratio of the angular diameter distances between
the deflector and the source and between the observer and the source.
We get a sensible value of $\sigma \lesssim$ 350\,\kms\  only if
$z_{\rm lens} \lesssim 1$. Furthermore, from the Faber-Jackson relation
(e.g., Faber et al.\ 1987; Djorgovski \& Davis 1987),
one can estimate the rest-frame luminosity of such a deflector and,
for a SED typical of a massive galaxy, one can predict the redshifted
$r$-band magnitude. For any redshift  $z_{\rm lens} \lesssim 1$,
the deflector would have a magnitude $\leq 24$ and be bright enough
to be detectable in the {\it HST} image. We thus conclude that, unless the
deflector is very atypical relative to more local lensing galaxies
(e.g., Auger et al.\ 2010), one would be able to detect it.

The brightest near-IR object in the vicinity of BR1202--0725
is the faint Ly$\alpha$~1 companion galaxy located between
the NW and SE sources, which
has an $r$ magnitude of 24.3.
It cannot play the role of a deflector, as it has
the same redshift as BR1202--0725 and is much too faint.  The second, even
fainter, Ly$\alpha$~2 companion galaxy lies 3$''$ southwest of
BR1202--0725, but it is again at the same redshift as the
quasar (Hu et al.\ 1996; 1997).

Finally, there is no arc-like feature on deep optical images that
would indicate very strong magnification of the bright quasar.
Such an arc-like feature might arise if
the quasar (only) were on the infinite-magnification caustic of
a lens so close to the line of sight to the quasar itself,
 that the lensing galaxy would be visible neither against the quasar light,
nor even on images where the quasar light is subtracted out, such as the
quasar-subtracted images of Fontana et al.\ (1996) and Hu et al.\ (1996).

In summary, there is no evidence, at any wavelength, for a galaxy-mass lens
along the line of sight to BR1202-0725, that could produce a submm double image.

\subsection{Absence of a nearby cluster lens for overall magnification}
Apart from trying to explain the submm double source by
lensing by a galaxy-mass deflector, one might ask whether the entire
BR1202-0725 complex could be lensed by  foreground cluster of galaxies.
In this case, the NW and SE submm sources are indeed independent galaxies,
but each of them, and the optical quasar, might be amplified by a comparable amount,
by a large, low-redshift foreground lens, i.e., a cluster of galaxies.
Here, the answer is no.   The BR1202-0725 region has been the object of a deep-field
survey, the NTT BR1202 field, together with an overlapping region, the NTT Deep Field,
resulting in a total field of $2.3'\times 4'$ (Fontana et al.\ 2000).
These two overlapping deep fields were surveyed in {\it UBVr} bands and in the near-IR {\it IJHK} bands.
Magnitude limits were of order 26 in {\it r} and 21.7 in {\it K}.  Although there is a peak in the
redshift distribution at $z\sim 0.6$ (Giallongo et al.\ 1998), there is no evidence for a cluster
of galaxies that could act as an overall magnifier for the BR1202-0725 complex as a whole.
Nor is there any systematic lengthening of galaxy images into ellipses or arcs that would
be signatures of large-scale magnifcation across the field.

\clearpage

\end{document}